\documentclass[11pt]{article}

\usepackage[inline]{enumitem}
\usepackage{amsmath}
\usepackage{amsthm}
\usepackage{xspace}
\usepackage{caption}

\usepackage[breaklinks=true,
            hidelinks,
            hyperindex,
            colorlinks = true,
            linkcolor = blue,
            urlcolor  = blue,
            citecolor = blue,
            anchorcolor = blue]{hyperref}
\usepackage{breakurl}
            
\newcommand{\mailto}[1]{\href{mailto:#1}{\nolinkurl{#1}}}
\usepackage{color}
\usepackage{soul}
  \definecolor{lightblue}{rgb}{.60,.60,1}

\theoremstyle{definition}
\newtheorem*{example}{Example}
\newtheorem{idea}{Idea}

\theoremstyle{remark}
\newtheorem*{remark}{Remark}
\title{A scalable verification solution for blockchains}

\author{
Jason Teutsch\\
\emph{TrueBit Establishment}\\
\mailto{jt@truebit.io}
\and 
Christian Reitwie{\ss}ner\\
\emph{Ethereum Foundation}\\
\mailto{chris@ethereum.org}
}

\date{November 16, 2017}
\newcommand{\truebit}{\textsf{TrueBit}\xspace}
\newcommand{\task}{\texttt{task}\xspace}

\newcommand{\reward}{\texttt{reward}\xspace}
\newcommand{\mindeposit}{\texttt{minDeposit}\xspace}

\newcommand{\timeout}{\texttt{timeOut}\xspace}
\newcommand{\solver}{\textsf{Solver}\xspace}
\newcommand{\verifier}{\textsf{Verifier}\xspace}

\newcommand{\solution}{\texttt{solution}\xspace}

\numberwithin{equation}{section}

\begin{document}

\maketitle

\begin{abstract}
Bitcoin and Ethereum, whose miners arguably collectively comprise the most powerful computational resource in the history of mankind, offer no more power for processing and verifying transactions than a typical smart phone.  The system described herein bypasses this bottleneck and brings scalable computation to Ethereum.  Our new system consists of a financial incentive layer atop a  dispute resolution layer where the latter takes form of a versatile ``verification game.''  In addition to secure outsourced computation, immediate applications include decentralized mining pools whose operator is an Ethereum smart contract, a cryptocurrency with scalable transaction throughput, and a trustless means for transferring currency between disjoint cryptocurrency systems.
\end{abstract}

\tableofcontents

\section{Securing computations with economics} \label{sec:economics}

Every Nakamoto consensus-based cryptocurrency (e.g.\ Bitcoin or Ethereum) offers something that everyone in the world agrees on: a definitive, public ledger of financial transactions, or \emph{blockchain}.  This consensus technology enables basic Bitcoin transactions, which transfer currency from one party to another, while Ethereum transactions perform financial and database operations contingent on the evaluation of more complex computational scripts.  Anonymous miners, who freely join and leave cryptocurrency networks, determine the validity of these transactions and thereby establish who owns which coins.  Remarkably, this verification scheme requires no central authority to operate securely.

In practice, miners can successfully maintain blockchain integrity so long as the computational burden of verification remains minimal.  Indeed, Bitcoin and Ethereum, whose miners arguably collectively comprise the most powerful computational resource in the history of mankind, offer no more power for processing and verifying transactions than a typical smart phone.  One cannot simply increase the volume or complexity of transactions flowing into the blockchain without risking inclusion of invalid transactions due to the so-called \emph{Verifier's Dilemma}~\cite{LTKS15} (see Section~\ref{sec:verifiersdilemma}).   In 2015, the Verifier's Dilemma manifested itself in the form of the July~4 Bitcoin fork~\cite{july4fork} which temporarily derailed the entire network.  The Ethereum community also witnessed a related exploit in its 2016 denial-of-service attacks~\cite{ethdos1}.

A \emph{consensus computer}~\cite{LTKS15} permits users to outsource computations to the Ethereum network and receive \emph{correct} answers in exchange for payment so long as the effort required to verify solutions does not exceed the threshold induced by the Verifier's Dilemma.  Thus the trustless consensus computer offers a small but reliable kernel of semantic truth. 

\paragraph{Our contribution.}
We present here a system, called \emph{\truebit}, which amplifies the consensus computer's capabilities.  \truebit enables trustless smart contracts, in theory, to securely perform any computation task.  Moreover, \truebit vastly reduces the number of redundant network node computations used in traditional Ethereum smart contracts.  Presently every Ethereum miner has to independently replicate each smart contract action in its entirety, whereas \truebit outsources most computation work to a handful of entities.  In this way, \truebit makes secure computing affordable.

\subsection{Outsourced computation}
Let us ponder for a moment why cryptocurrencies might offer an especially convenient framework for secure outsourced computation, the core application of \truebit.  In the traditional cloud models, users must trust that the machine hardware, software, and cloud administrator all perform as expected.  A wide range of things can go wrong, particularly when one wishes to tie the results of such computations to monetized entities such as smart contracts.  Proper economic incentives, the cornerstone of any cryptocurrency, can deter many types of errors from occurring in ways that simple task repetition cannot.  Furthermore, in contrast to a cloud whose configuration details may not be visible to users, any systemic network errors that might occur on a blockchain-based system like \truebit would appear in plain sight to the entire community.  Cryptocurrencies indeed provide an excellent starting point as they already achieve several desirable properties.
\begin{enumerate}
\item As witnessed by the Ethereum consensus computer, Nakamoto consensus grants networks the ability to trustlessly perform small computations correctly.

\item Blockchain public ledgers provide perfect transparency and immutability.  Ethereum smart contracts inherit these characteristics.

\item Currency incentives, which in cryptocurrencies are inextricably tied to computational processes, can be used to recruit and reward participants.
\end{enumerate}
In general, economic forces greatly reduce the scope of possible network errors.  We can safely assume that errors which incur economic penalties are much less likely to occur than those which do not.

\subsection{Practical impact}
A market for work related to computationally-intensive smart contracts already exists.  The Golem Project, which crowdfunded 820,000 ETH on the morning of November~11, 2016 ($\sim$\$8.6 million on that date), already cites \truebit as a verification mechanism for their forthcoming outsourced computation network \cite{golemwhite}.  \truebit has the potential to support many practical new applications.  Here are some examples.
\begin{itemize}
\item \textit{Outsourced computation.} \truebit functions as a worldwide computation market.  Anyone in the world can post a computational task, and anyone else can receive a reward for completing it.  The system's incentive structure guarantees correctness of returned solutions.

\item \textit{Truly decentralized mining.} Centralized mining pools pose security threats.  For any Nakamoto consensus-based cryptocurrency, one can use \truebit to build a practical, efficient, and trustless mining pool managed by an Ethereum smart contract (see also \cite{LWTS16re}).

\item \textit{Trustless cryptocurrency exchange.} Currently users must find trusted exchange partners to transfer currency (or messages) between blockchains.  \truebit can facilitates secure transfer, for example, of dogecoins to the Ethereum blockchain and back (modulo some new opcodes for dogecoin) \cite{dogether}.

\item \textit{Scalable blockchain.} By decoupling verification for miners into a separate protocol, we can achieve high transaction throughput without facing a Verifier's Dilemma.

\item\textit{Scalable ``on-chain'' storage.} Swarm \cite{swarm} is developing a platform for incentivized, peer-to-peer storage.  \truebit can make Swarm data accessible to Ethereum smart contracts.
\end{itemize}
We discuss these ideas further in Section~\ref{sec:applications}.

\subsection{Smart contracts}
\emph{Smart contracts}, introduced in 1994 by Szabo~\cite{nick-szabo} and first realized in the Ethereum~\cite{ethereum} cryptocurrency in 2015, are uncensorable programs that live on Ethereum's blockchain and have their own executable code and internal states, including storage for variable values and ether currency balance.  Proposed applications for smart contracts include outsourced computation and storage~\cite{golem,swarm,LTKS15}, prediction markets~\cite{augur}, decentralized venture funds~\cite{thedao}, and Bitcoin mining pools~\cite{smartpool,LWTS16re}.  Ethereum's variable \texttt{gasLimit} restricts the number of computation steps that smart contracts can perform.  

\truebit itself is an Ethereum smart contract which allows users to call \emph{\truebit contracts} for trusted, computationally-intensive applications.  Traditional Ethereum smart contracts can call a \truebit contract as a subroutine, thereby effectively bypassing Ethereum's \texttt{gasLimit}.  In other words, \truebit increases the per block work that the Ethereum network can process correctly.

\section{How \truebit works} \label{sec:htw}
\truebit's primary purpose is to realize correct, trustless computations despite miners' limited computation bandwidth.  Intuitively, we wish to reward participants who correctly perform computational tasks, but who decides whether these tasks were done correctly?  In absence of a dispute, the party who performs a computational task on behalf of a \truebit contract simply receives a reward.  On the other hand, if a dispute does occur, we must rely on the only trusted resource, the limited network of miners, to resolve it.  Dispute resolution occurs as a ``verification game'' subroutine in \truebit, and we will discuss its details in Section~\ref{sec:amplification layer}.

Who initiates a dispute in the case of an incorrect computation, and how can we ensure that such a dispute will actually occur?  One option is to assume that each party with some stake in the contract brings their own trusted (although not necessarily mutually trusted) verifier, and flags for a challenge if needed.  This trusted verifier approach suffices for some applications, but in general parties may need to rely on the results of a previously executed \truebit contract \emph{ex post facto} even if they themselves did not exist at the time of its creation.  Like Ethereum smart contracts, \truebit contracts must have universal validity.

Since there exist no trusted parties on Ethereum's network, by symmetry we must allow any party to be hired to solve any computational task, and similarly anyone should be able to challenge a Solver's outcome of a computational task.  The latter requirement ensures that \truebit operates by unanimous consensus.  Assuming that an honest and motivated Challenger exists who will challenge any wrong answer given by the Solver, the system's incentives for the Solver are straightforward: reward correct solutions.  But how can we make sure that such a motivated Challenger exists?
\begin{idea}
Offer rewards for checking computations.
\end{idea}
While this strategy may encourage participation, it provides no incentive for correct verification.  We have no \emph{a priori} reason to trust that the Verifier will substantively inspect the computation task, and we gain no additional assurance by increasing her reward for checking.
\begin{idea}
Offer rewards for finding bugs.
\end{idea}
While a bug bounty may incentivize a Verifier to correctly perform a check, it only does so when the Verifier has some reason to believe that a bug might actually exist.  Unless a potential Verifier believes that she has some real chance to find a bug, in practice we cannot expect her to participate in the system.  In summary, we need both an incentive to partake in verification and an incentive to perform checks correctly.  This leads us to the following proposition.
\begin{idea}
Offer a bug bounty \emph{and} provide expectation that bugs will exist.
\end{idea}
\truebit takes this third approach by occasionally forcing Solvers to submit incorrect solutions.  During ``forced errors'' \truebit reverses the normal system incentives: the Solver gets paid for submitting an incorrect solution and penalized for submitting a correct one.  We discuss the details of \truebit's forced error protocol in Section~\ref{sec:incentive layer}.

Let us return to our dispute resolution subroutine.  At the heart of \truebit's protocol lies an interactive ``verification game'' which decides whether or not a contested computational task was performed correctly (see Section~\ref{sec:amplification layer}).  The verification game proceeds through a series of rounds, where each round recursively checks a smaller and smaller subset of the computation.  A trusted network, in our case the Ethereum platform \cite{ethereum}, merely enforces the rules of the game and therefore does not bear the bulk of the verification burden.  Anyone on the network can generate tasks, compute, or verify in exchange for rewards.  \truebit does not require ``honest'' or ``altruistic'' nodes for correct operation but rather runs under the assumption that each node wishes to maximize its own profit.  We shall discuss the details of \truebit's incentives for participation in Section~\ref{sec:incentive layer}.

\subsection{System properties}

Any outsourced computation system should be \emph{fair} in the sense that parties who perform computational tasks indeed receive compensation for their work and \emph{reliable} in the sense that paid computations are performed correctly.  In addition to these properties, \truebit also ensures \emph{accessibility} through easy joining or exiting of the verification ecosystem.  Any \emph{anonymous} party can play the role of Task Giver, Solver, or Verifier, where a Verifier is a party who checks solutions and becomes a Challenger whenever she reports an error.  In particular, \truebit does not trust or rely on the reputation of its participants.  Everyone who puts a deposit into the system has a fair chance to be hired for a given computational task.

\truebit offers several novel advantages over traditional cloud computing and verifiable computable models.  Verifiable computing ensures a correct answer for an outsourced computation by forcing the cloud to provide a short proof which witnesses the correctness of its computation.  The idea is that this ``short proof'' should be much easier to check than performing the computation oneself.  Researchers have achieved much progress on this method in recent years, however the cryptographic setup costs and computational overhead for the cloud in state-of-the-art systems make these methods unsuitable for most real-world applications.  Moreover, many of the proof-based systems to-date, including Zaatar, Pinocchio, Ginger, and TinyRAM, require one to run thousands of instances of a single function before breaking even on the overall verification time for a $128 \times 128$ matrix multiplication puzzle \cite{WB15}.  The new cryptocurrency Zcash \cite{zcash} successfully makes use of verifiable computing machinery, albeit only for a very restricted class of computation tasks.  Below we contrast \truebit with verifiable computing and traditional cloud computing.
\begin{enumerate}
\item \emph{Incentives.} Unlike traditional cloud computing, where the user simply trusts the cloud to provide correct answers, \truebit provides financial incentives to ensure correctness.

\item \emph{Transparency.}  The entire inner workings of \truebit's interpreter sit on the blockchain and are open for inspection (see Section \ref{sec:implementation}).  Furthermore, the user community can democratically update the interpreter as needed.

\item \emph{Efficiency.}  Solvers in \truebit have low computational overhead and minimal initial setup costs.  The verification game (Section~\ref{sec:amplification layer}) does introduce some extra work, but in practice, due to high penalties for wrong answers and bogus challenges, we expect participants to appeal to the verification game only rarely, if at all.

\item \emph{Simplicity.}   \truebit's operation is relatively straightforward.  Unlike traditional verifiable computing, \truebit avoids deep probabilistically checkable proofs (PCPs), succinct non-interactive arguments of knowledge (SNARKs) \cite{BCGTV13}, and exotic cryptographic assumptions (e.g.\ those used in zkSNARKs \cite{BCPR13}).  The standard cryptography used in \truebit, namely hash functions and digital signatures, sits in the underlying blockchain network and does not surface in the \truebit protocol itself.

\item \emph{Adaptability.} \truebit runs on top of Ethereum's current protocol without impacting functionality.

\item \emph{Keyless entry.}  Participants do not need to manage cryptographic keys beyond those used in Ethereum.  \truebit establishes identities through financial deposits alone, and therefore the system avoids risks from cryptographic trapdoors.
\end{enumerate}

\subsection{Assumptions} \label{sec:netass}
Traditional distributed systems models focus on tolerance against arbitrary, Byzantine errors.  In Ethereum  and other Nakamoto consensus-based cryptocurrencies, however, we have no reason to expect arbitrary errors---nodes generally deviate from protocol due to financial incentives.  Our system model makes two basic assumptions about the underlying network.
\begin{enumerate}[label=\textsc{(\roman*)}]
\item There exists a trusted network (i.e.\ Ethereum) that correctly performs very small computation tasks.

\item Participants are rational in the sense that they act to maximize their individual profits.  In particular, we expect that CPUs will show up to do computational tasks if and only if they expect fair compensation for their work.
\end{enumerate}
The consensus computer~\cite{LTKS15} shows how one can use the incentives in Ethereum to establish assumption~\textsc{(i)} above both in theory and in practice.  The task of our new system will be to amplify this small amount of correct computation in order to handle larger tasks using assumption~\textsc{(ii)}.  Even though solo Ethereum miners rarely earn mining rewards, expected long-term returns suffice to merit their participation in the system (via mining pools).  As we shall discuss in Section~\ref{sec:verifiersdilemma}, anonymous parties may not necessarily perform correct computations when economic incentives, including conservation of CPU cycles, pressure them to do otherwise.  This observation justifies assumption~\textsc{(ii)}.  We emphasize that we do not find the assumption that there exists a single honest (and non-lazy) participant \cite{JSST16} to be realistic.  A person who loyally contributes CPU cycles without incentive to do so most likely does not exist in a large, general purpose, public system.

This paper takes a somewhat simplified view of Nakamoto consensus.  In some cryptocurrencies, such as Bitcoin, miners can selectively censor a class of transactions by deciding not to include its members in their blocks.  In the case of \truebit, censorship of challenges (see Section~\ref{sec:incentive layer}) or transactions in the verification game (Section~\ref{sec:verificationgame}) could result in catastrophic, bogus solutions being accepted on the blockchain.  In Ethereum, however, censoring individual transactions is difficult as miners cannot easily see where an Ethereum transaction might call without executing it.  Indeed Ethereum users can obfuscate the function of their transactions, making it computationally costly for miners to hand-pick transactions for their blocks.  Hence we shall assume that censorship of individual transactions does not happen.  In Section~\ref{sec:sybil}, we analyze an attack on \truebit in which miners withhold entire blocks and show that it is not profitable.

\subsection{Attacker model} \label{sec:attackermodel}
\truebit's security relies on two basic assumptions about the behavior and capabilities of potential attackers.
\begin{enumerate}[label=\textsc{(\roman*)}]
\item \emph{Attackers cannot subvert the underlying consensus computer.}  The Judges in the verification game (Section~\ref{sec:amplification layer}) always adjudicate correctly, as do the Referees in the incentive layer (Section~\ref{sec:incentive layer}).

\item \emph{Attackers are rational.}  We assume that attackers have limited financial resources and will not deviate from protocol unless it is profitable for them to do so.
\end{enumerate}
 While an adversary with sufficient computational power can derail a Nakamoto consensus-based cryptocurrency \cite{Nak09}, assumption~\textsc{(i)} above neatly \linebreak sweeps this issue outside our domain.  \truebit itself does not utilize proof-of-work.  Although we have chosen to implement \truebit in the Ethereum platform for convenience, we could easily replace the ``Judges'' from~\textsc{(i)} with any other universally trusted computing mechanism without affecting the system's functionality.

Due to anonymity of all participants, \emph{Sybil attacks}, in which a single party deviantly assumes multiple identities on the network, pose special threats to \truebit (see Section~\ref{sec:sybil} and Section~\ref{sec:trifecta}).  We will assume that an attacker may create as many identities as she likes, including simultaneously playing the roles of Task Giver, Solver, and Verifier, appearing as two distinct Verifiers, or managing multiple identities via pooled financial resources (see Section~\ref{sec:deeppocket}).  Using assumption~\textsc{(ii)} above, we shall argue that \truebit resists such attacks in Section~\ref{sec:defenses}.

The ``forced errors'' described in the beginning of this section (Section~\ref{sec:htw}) pose a special challenge to \truebit due to opportunistic attacks.  As the prize for discovering forced errors is necessarily enormous, attackers may seek ways to obtain these large prizes prizes without performing the intended verification tasks (Section~\ref{sec:lhfruit}).  For this reason, the occurrence of forced errors must appear truly random to all prospective Verifiers.  Moreover, Solvers must be incentivized from prematurely sharing the fact that their error was forced or claiming that an error was forced when it really wasn't.  In general, we want consistent, not sporadic, verification.  At least one Verifier should be present for every task.  The attraction of forced errors represents the ``Achilles' heel'' of \truebit, and we devote the incentive structure in Section~\ref{sec:incentive layer} to careful defending of this potential attack surface.

\section{Dispute resolution layer} \label{sec:amplification layer}

In this section we present and analyze \truebit's dispute resolution mechanism.  \truebit relies on network ``Judges'' with limited computational power who adjudicate all disputes, and we show how such Judges can correctly resolve disputes over complex computations.

\subsection{Bottleneck: The Verifier's Dilemma} \label{sec:verifiersdilemma}

Ethereum's incentive structure severely restricts the amount of computation work that smart contracts can accurately enforce.  Users interact with smart contracts by broadcasting \emph{transactions} to the network.  Such transactions contain program scripts which require computational work to verify their validity prior to miners' acceptance on the blockchain. Each miner must not only locally verify but also remember the state of every single smart contract.  On the surface, this redundant work limits smart contracts' computational scope due to inefficiency, but the true bottleneck lies deeper at the incentive level.

Let us recall miners' incentives for participating in Ethereum.  In Nakamoto consensus-based cryptocurrencies~\cite{Nak09}, new transactions enter the blockchain through a process called \emph{mining}.  Participating \emph{miners} maintain the integrity of the blockchain data by racing to solve computation-intensive, \emph{proof-of-work} puzzles  in exchange for rewards. The first miner who successfully broadcasts a solution to the current proof-of-work puzzle proves that she has spent the necessary computation power to merit appending her new set of transactions to the blockchain, and this step awards the miner a set of newly minted coins.  Then the race begins again on top of this new block.

When an Ethereum user broadcasts a transaction to the network, the miner who appends her valid transaction to the blockchain receives a \emph{transaction fee} for the service.  On the other hand, Ethereum's mining protocol dictates that other miners should verify this transaction \emph{gratis}, for the ``common good'' so to speak.  More specifically, the mining protocol dictates that the longest chain consisting of valid transactions is the correct one, and miners should mine on top of it.  If the time to verify new blocks is negligible, miners may indeed agree to check them, but if the verification work becomes substantial, then miners risk falling behind in the mining race by following protocol and actually verifying.  In order to save time, rational miners might simply skip the verification step, leaving the blockchain open to invalid transactions.  On the other hand, by skipping verification, a miner might also waste her CPU cycles by mining on top of a stale chain which the rest of the community deems invalid.  Either way, the miner risks forfeiting rewards.  A rational miner doesn't know what to do!

Any substantial verification burden placed on miners thus results in a \emph{Verifier's Dilemma}~\cite{LTKS15}.  Because of the Verifier's Dilemma, smart contracts whose verifications require non-trivial computational efforts will fail to execute correctly as rational miners, who may choose to deviate from the suggested protocol, stop donating their limited resources to thankless verification tasks.  In short, Ethereum lacks a scalable verification mechanism and remains vulnerable to serious attacks \cite{LTKS15} when miners have to verify more than a tiny amount.

\truebit contracts securely execute computationally-intensive tasks for use in Ethereum smart contracts.  The system's core consists of a smart contract interface where a user can request a solution to a computational task or puzzle, and anyone can solve it in exchange for a reward.  Our interactive verification game, described below, empowers users to both outsource arbitrary computations in a decentralized manner and receive back correct solutions.  To achieve this goal, we will not only reduce the redundancy of verification work but also refine the incentive structure responsible for the Verifier's Dilemma.

\subsection{Solution: The verification game} \label{sec:verificationgame}

The goal of \truebit's interactive verification game is to resolve a given dispute between a Solver, who has provided a solution to some computational task, and a Challenger, who disagrees with the solution and therefore calls the verification game to be played.  The outer layer of \truebit (Section~\ref{sec:incentive layer}) uses the verification game as a subroutine.  The roles in the verification game include a Solver, who offers a solution to a given task, and a Challenger who disagrees with the Solver's solution.  The final party, the Judges, always perform computations correctly but possess extremely limited computational bandwidth.  The Judges in \truebit are the entire community of Ethereum miners who reach verdicts through Nakamoto consensus.

The verification game proceeds in a series of rounds, each of which narrows down the portion of the computation in dispute.  In the first round, the Challenger forces the Solver to commit to a pair of computation steps delimiting some time interval.  In the next round, the Challenger iteratively challenges a subset of the Solver's computation over this time interval, and next she challengers over a subset of that subset, etc.\ until in the final round the final challenge is sufficiently trivial that the Judges can make a final ruling on whether the challenge was justified.  The Judges also enforce that the Solver and Challenger follow the rules of the game.  At the end of the verification game, either the cheating Solver will be discovered and punished in the outer layer of \truebit (Section~\ref{sec:incentive layer}), or the Challenger will pay for the resources consumed by the false alarm.

\begin{example}[matrix multiplication \cite{JSST16}]
We give an illustrative example of a verification game.  Suppose that the given task is to compute the product of two matrices $A$ and $B$.  The Solver claims that $A \times B = C$.  In the first round, the Challenger must claim an error in one of the entries in $C$, and outputs coordinates $i,j$ such that
\[
c_{i,j} \neq \sum_{k=1}^n a_{i,k} \cdot b_{k,j}
\]
with corresponding evidence consisting of partial sums $d_0,d_1,\ldots,d_n$ where 
\[
d_m = \sum_{k=1}^m a_{i,k} \cdot b_{k,j}.
\]
The Judges only verify that $i,j$ are coordinates and that $d_n \neq c_{i,j}$ and that $d_0 = 0$. The Challenger loses if this
fails to hold.  In the second round, the Solver tries to defend himself by providing a $k$ such that $d_k \neq d_{k-1} + a_{i,k} \cdot b_{k,j}$. If the Judges can verify this claim then the Solver wins the game, and otherwise the Challenger wins.
\end{example}

\subsection{Detailed protocol} \label{sec:detailedprotcol}

We can use a variation of the Example above to check arbitrary computations by creating a binary search to pinpoint the erroneous (or not erroneous) step in a computation.  The idea for the following verification game is essentially due to Canetti, Riva, and Rothblum \cite{CRR11,CRR13} and was later independently discovered by the authors of this paper \cite{JSST16,notsmartjudges}.  Canetti, Riva, and Rothblum did not consider this game in the context of blockchains but rather a simpler scenario in which a user outsources a computation to $k$ different clouds under the assumption that at least 1 out of $k$ of the clouds will perform the computation correctly.  The assumption of at least one trusted Solver/Challenger cloud is too strong for our purposes since in a purely rational network, such a Solver/Challenger may not exist.  Section~\ref{sec:incentive layer} describes our workaround for the smart contract setting.

For purposes of reducing the number of rounds of interaction and space demands on the blockchain, our protocol roughly follows \cite{CRR11,CRR13} which combines the (parametrized) binary search optimization from \cite[Proposition~9]{JSST16} with the probabilistic Merkle tree construction from \cite{notsmartjudges}.  Recall that a \emph{Merkle tree} is a binary tree in which each node is the hash of the concatenation of its children nodes.  In general, the leaves of a Merkle tree will collectively contain some data of interest, and the root is a single hash value which acts as a certificate commitment for the leaf values in the following sense.  If one knows only the root of a Merkle tree and wants to confirm that some data $x$ sits at one of the leaves, the holder of the original data can provide a path from the root to the leaf containing $x$ together with the children of each node traversed in the Merkle tree.  Such a path is difficult to fake because one needs to know the child preimages for each hash in the path, so with high likelihood the data holder will supply a correct path if and only if $x$ actually sits at the designated leaf.

For clarity, we will present the verification game in the familiar and standard language of Turing machines, but in actuality \truebit uses the Google Lanai architecture to simulate steps in the computation (see Section~\ref{sec:implementation}).  The verification game consists of three parties:
\begin{itemize}
\item  a \emph{Solver} who has allegedly performed some computational task and claims some solution,
\item  a \emph{Challenger} who disputes the correctness of the Solver's solution, and 
\item \emph{Judges} with bounded computational power, who will rule on whether the output is correct.
\end{itemize}
\truebit fills the roles of Solver and Challenger according to the incentive layer discussed in Section~\ref{sec:incentive layer}.  The Judges are the set of miners in Ethereum who collectively record their decisions on the blockchain.  The incentive layer (Section~\ref{sec:incentive layer}) determines the financial consequences of the Judges' verdict.

To start, the Solver and Challenger each privately compile a tableau of Turing configurations mapping each time step of the task to its full internal state.  Assume that the task runs in time $t$ steps and space $s$, where $s$ bits suffice to include the complete state of the Turing machine at any given time, including tape contents, head position, and machine state.  The game protocol includes a fixed, integer parameter $c>1$ which says how many configurations the Solver broadcasts to the blockchain in each round of the verification game.  More configurations per round mean fewer total rounds needed, however it also means that the total number of configurations sent to the blockchain increases.  In particular, we reach diminishing returns as the data sent per round reaches Ethereum's per block capacity.

The verification game protocol proceeds as follows.  
\begin{description}
\item[Main loop.]
The following steps are done iteratively to pinpoint the source of disagreement.  The protocol determines timeout periods within which Solvers and Challengers must respond.  Failure to respond within time bounds results in immediate loss for the non-responding party.
\begin{enumerate}
\item The Solver selects $c$  configurations equally spaced apart in time across the current range of the dispute.  In the initial iteration, for example, the Solver selects configurations across the entire tableau at the following time steps:
\[
\frac{t}{c}, \frac{2t}{c}, \frac{3t}{c}, \dots, \frac{ct}{c}.
\]
He then produces $c$~Merkle trees, each with $s$ leaves, where the leaves constitute the complete machine state at each of these times, and broadcasts each of the roots of these Merkle trees to the blockchain. 

\item The Challenger responds with a number~$i \leq c$, indicating the first time step in this list that differs from her own, and broadcasts this number~$i$ to the blockchain.

\item The Judges check that the Solver indeed posted $c$ Merkle roots to the blockchain and that the Challenger's value~$i$ satisfies $1 \leq i \leq c$.  If either of these checks fails, the Solver, or respectively the Challenger, loses immediately.

\item The next iteration of the game continues similarly, but restricted to configurations between the $i-1$-st and $i$-th indexed configurations.  Here we interpret a $0$ as the computation's initial configuration.
\end{enumerate}

\item[Final stage.]  After $\log t / \log c$ rounds, the loop above converges to the first, disputed computational step, and the Judges explicitly check this step of the computation as follows.  Suppose that the first disagreement occurs at time~$e$.  The Solver provides paths from the Merkle root for time~$e$ to its leaves containing:
\begin{itemize}
\item the location of the machine head at time~$e$,

\item the content of the tape square under the machine head, the tape contents of its left and right neighbors, and

\item the machine state at time~$e$.
\end{itemize}
The Solver also provides the analogous paths for time~$e-1$.  The Solver loses the dispute if the Judges find that any of these paths fail to be valid.  Finally, using this information and the original task code (which existed on the blockchain prior to the verification game), the Judges check the computational step at time~$e$ and rule whether the challenge was justified.
\end{description}

\subsection{Runtime and security analysis} \label{sec:vgruntime}

We now show that the work done by the Judges is small compared to the work required to perform the task, and that the number of interaction rounds is modest.  Note that the majority of effort from the Judges simply involves recording data onto the blockchain.  The Judges' work is split over many transactions and therefore avoids any chance of a Verifier's Dilemma.

We fix $\sigma$ as the size of each hash in the Merkle trees, and $p$ as the space required to store a single machine state for the computation.  For computational tasks that run in time $t$ and space $s$ with $c$ configuration roots posted per round, the protocol requires $\log t/\log c$ rounds of interaction on the main loop, writes $c \sigma + \log c$ bits to the blockchain during each iteration of the main loop, and writes $2 \cdot ((2\sigma+1) \log s + 5 + p)$ bits in the final round (assuming that tape squares can contain either 0, 1, or be blank, so we need $3 \cdot \log 3/\log 2 \approx 5$ bits to describe 3 of them).  Summing these parts, the total time required for community-wide verification from the Judges is then
\[
O \left[ \frac{\log t}{\log c} \cdot \left( c \sigma + \log c \right) + 2 \cdot ( (2\sigma+1) \log s + 5 + p) \right]
\]
where the hidden constant depends on the implementation of this protocol on the Judges' local machines.  This estimate does not include the space and time required to store the instructions for the initial computation task.  We assume that this data already exists on the blockchain before the verification game begins.

In comparison with other proposed blockchain-based computation systems, or even PCP-based verifiable computing systems \cite{WB15}, the work imposed on the Judges by the verification game is small \cite{computationmarket}, as is the computational overhead required to join the system~\cite{HKZ14}.  Moreover, the economic costs to the Task Giver, who pays only a Solver and a Challenger and not all Judges to process the full task, is modest, as are the computational overheads for the Solver and Challenger.

Note that larger values for $\sigma$ reduce the chances of hash collisions.  Hash collisions might allow a dishonest Solver to substitute configurations in his computation tableau with fake ones sharing the same Merkle roots.  Two consecutive configurations in the tableau whose Merkle roots happen to agree with the Merkle roots of two consecutive configurations that the dishonest Solver wishes to apply as substitute suffice to disrupt the integrity of the verification game protocol.  Indeed, the Challenger will agree with both Merkle roots, and the Judges will confirm that the transition between them is valid and therefore incorrectly reject the challenge.  For a fixed size $\sigma$, the chance that such collisions occurs by accident even becomes likely for sufficiently enormous tasks.

One could entirely eliminate the security risk discussed in the previous paragraph by posting complete machine states on the blockchain rather than just Merkle roots, but this makes the protocol more expensive.  Alternatively, one could either increase the parameter $\sigma$ or the number of states checked by the Judges in the final step.  In this sense,  the choice of $\sigma$ bounds the maximum complexity of secure computations in \truebit.  We will also see in Section~\ref{sec:jackpot} that the capital held by the \truebit contract as a ``jackpot'' for finding a forced error poses an additional constraint on computational capacity, but this value can scale as well.  Finally, the effectiveness of \truebit's verification game may degrade for extremely complex tasks due to the computational limitations of the Challenger.  If the expected execution time of a verification game exceeds a human life expectancy, for example, then potential Challengers may lack motivation to participate despite financial incentives.  Note that even when a task itself is feasible, its corresponding verification game may not be as the verification game carries significant overhead.  We anticipate that future versions of \truebit may optimize the verification game for certain kinds of tasks so as to reduce this discrepancy (see Section~\ref{sec:bigdata}).  

\begin{remark}
The verification game provides a significant privacy-protecting effect in that only disputed parts of a given computation must touch the public eye~\cite{prubyreddit}.  
\end{remark}

\section{Incentive layer} \label{sec:incentive layer}
We now discuss the financial incentives which encourage anonymous parties to both contribute CPU cycles and perform requested computational tasks correctly.  We use the dispute resolution layer from Section~\ref{sec:amplification layer} as a subroutine and connect its role to the present construction.  A \emph{Verifier} is a potential Challenger, that is, someone who checks submitted solutions and calls for a challenge (see Section~\ref{sec:amplification layer}) if and when she detects an error.  By assumption~\textsc{(ii)} in Section~\ref{sec:netass}, Verifiers must receive adequate payment for verification tasks.  As argued in the beginning of Section~\ref{sec:htw}, however, Verifiers can only receive rewards when they actually find bugs.  It follows that the reward for discovering errors must encompass amortized payment for all verification tasks checked, including those in which no bugs were found.

Any Ethereum user who offers a reward through \truebit for performing a computational \emph{task} is called a \emph{Task Giver}.  A party which offers a \emph{solution} for performing these tasks in exchange for a reward is called a \emph{Solver}, and, as indicated in the previous paragraph, Verifiers check that Solvers' solutions are correct.  Solvers and Verifiers will play the roles of ``Solvers" and ``Challengers'' respectively from the dispute resolution layer (Section~\ref{sec:amplification layer}) whenever a Verifier initiates a verification game.  \emph{Referees} enforce the rules of the \truebit protocol in the incentive layer.  While Ethereum miners enforce protocol rules both through the roles of ``Judges'' in Section~\ref{sec:amplification layer} and the role of ``Referees'' in the present section, we use distinct labels for them because the layers and functions differ for these roles.   The primary role of Judges is to interactively resolve a dispute, whereas Referees primarily enforce that Solvers and Verifiers timely submit appropriate data in the incentive layer.

\paragraph{In a nutshell.}  The main steps of a \truebit contract are as follows.  Items in quotes will be throughly explained in due course.
\begin{enumerate}
\item A Task Giver announces a task and offers a reward for its solution.

\item A Solver is elected by lottery, and he prepares both a ``correct'' and an ``incorrect'' solution to the task. 
\begin{enumerate}
\item If a ``forced error'' is in effect, the Solver reveals the incorrect solution on the blockchain.
\item Otherwise the Solver reveals the correct one.
\end{enumerate}
\item Verifiers check the Solver's solution.  They win a large ``jackpot'' payout if both:
\begin{enumerate}
\item they correctly identify the solution as erroneous, and
\item a forced error was in effect.
\end{enumerate}
\item If no Verifier signals an error, then the system accepts the solution.  Otherwise acceptance depends on the outcome of a verification game.
\end{enumerate}
In the rest of this section, we discuss the security measures in \truebit which ensure that participants comply with the skeletal procedure outlined above.  A more complete description of the protocol appears in Section~\ref{sec:truebit protocol}.

\subsection{Jackpots} \label{sec:jackpot}
\truebit periodically imposes \emph{forced errors} in which a Solver must offer a wrong solution to a task (see Section~\ref{sec:htw}).  This ensures that Verifiers who diligently search for errors will eventually find them.  Accordingly, Verifiers who correctly report forced errors receive substantial \emph{jackpot} payouts.   By design, Verifiers cannot predict when forced errors will occur and therefore have incentive to check all tasks carefully.  Forced errors occur only rarely, and we expect Solvers to make only few, if any, other errors.  Rewards for identifying unforced errors may be modest compared to jackpot payouts, and we consider any such payments of these type incidental, rather than fundamental, to \truebit's secure incentive structure.

The jackpot payout effectively bounds the complexity of computation tasks that \truebit can perform securely.  The Verifier must, on average, receive a payment which fairly compensates her for the task at hand, which means that the jackpot payout should at least consist of fair compensation for the current task times the forced error rate.  In this way, the jackpot cap bounds the Verifier's expected per task compensation which, by assumption~(\textsc{ii}) in Section~\ref{sec:netass}, restricts Verifiers' available CPU cycles per task.

We fix a rate for forced errors among tasks.  This fraction should not be so low so as to discourage Verifier participation through excessively infrequent rewards, but neither should it be so high so as to run the risk of Referees bias (see Section~\ref{sec:sybil}).   We set forced errors to occur, on average, once every thousand tasks.

\subsection{Taxes} \label{sec:tax}
Given that a jackpot \emph{repository} must exist, we now describe the mechanism for funding it.  We assume that a generous philanthropist deposits some initial funds into the repository, but thereafter \truebit will be self-sustaining through \emph{taxes}.  Any Task Giver that calls a \truebit contract must pay not only the cost of computational work done by the Solver but also for the work done by the Verifier(s) (excluding unforced errors and bogus challenges), as well as the work done by Referees and Judges.  We refer to the latter two costs as the \emph{verification tax}.  We ensure that the jackpot repository never disappears entirely by placing a cap on the jackpot size.  To this end, we set the maximum jackpot payout for a forced error to be one third of the total repository size.

While a single, attentive Verifier suffices to ensure correctness and achieves ideal tax efficiency, in practice the verification tax requires a substantial cushion.  We estimate the necessary verification tax to be 500\% -- 5000\% of the cost of performing the given task.  As we shall see in Section~\ref{sec:trifecta}, there is a quantitative tradeoff between tax efficiency, Solver deposits, and security of computations, so Solvers could potentially absorb some of this tax burden by contributing higher deposits.  Our tax rate estimate incorporates Verifiers' incentives for participation and the fact that both the Task Giver and the Solver for a given task each have incentive to perform verification.  Participation from these parties may necessitate higher taxes because the total jackpot payoff decreases exponentially as the number of challenges increases (see Section~\ref{sec:deeppocket}). Expected jackpot payoffs must be sufficiently high to consistently attract at least one Verifier per task.  The required tax amounts may also depend on peculiarities of human behavior.  Indeed, we may have to pay Verifiers more than Solvers per CPU cycle because Verifier rewards have the human-undesirable property of being sporadic whereas Solvers always receive immediate rewards.  Thus the optimal tax rate must, at least in part, be determined experimentally.

\subsection{Deposits} \label{sec:deposit}
\truebit requires deposits from Solvers and Verifiers in order to thwart Sybil attacks (see Section~\ref{sec:sybil}) and incentivize correct computations.  We set these deposits to be more than the expected jackpot payout for a given task plus the cost of playing a verification game.  In particular, the deposits must be large enough to:
\begin{enumerate}
\item pay for the (expensive) cost of a verification game, including all rewards and penalties for Solver and Challengers and work done by Judges,

\item discourage Solvers and Verifiers from sacrificing deposits in order to obtain jackpots without performing verification (see Section~\ref{sec:sybil} and Section~\ref{sec:deeppocket}),

\item discourage Task Givers who collude with Solvers in effort to get bogus solutions onto the blockchain (see Section~\ref{sec:trifecta}),

\item refund taxes to the Task Giver in case the Solver causes an unforced error,

\item deter Solvers from randomly guessing solutions to obtain task rewards instead actually performing computations (as might be profitable for binary decision tasks with very high task rewards), and

\item deter external, temporal pathologies.
\end{enumerate}
Note that the currency used to pay Solvers and Verifiers need not be the same as the currency used to pay Judges and Referees, but for simplicity of presentation, we fix ether (ETH) as the unique underlying currency.

As an example of the second type, consider a situation where the Solver deposit is small (say 10~ETH) but the expected jackpot payout per task is high (say 1000~ETH).   An individual playing both the role of the Solver and Verifier could offer a bogus solution and then challenge his own answer, hypothetically netting, on average, $1000-10 = 990$~ETH without providing any useful service.  Such an action would degrade other Verifiers' incentive to participate.

As an example of the last case, if the outcome of a time-sensitive \truebit contract controlled a 1000~ETH payout but only required  a 10~ETH Solver deposit, then the party who stands to lose 1000~ETH from the \truebit contract could attempt to cause a delay by posing as a Solver to the \truebit contract and giving a bogus answer.  We leave it to the Task Giver to determine appropriate minimum deposit values for such specific situations, as such contextual determinations lie outside of the scope of \truebit itself.

\subsection{Generating forced errors} \label{sec:randombits}
In order to motivate verification of all tasks and to guard the jackpot repository against swindle, forced errors must appear unpredictably.  \truebit uses strings of random bits to determine whether or not a forced error occurs for a given task.  The system derives its unpredictability via the following properties.
\begin{enumerate}
\item The Task Giver does not know the random bits at the time that she announces a task.

\item The Solver does not know the random bits until after he has committed his solution.

\item Verifiers do not know the random bits until after they decide whether to challenge.
\end{enumerate}
The first property makes it difficult for a Task Giver to create a task designed to swindle the jackpot repository, and the second discourages Solvers from being lazy and only volunteering to solve tasks which have forced errors.  In order to satisfy these three properties, \truebit combines random bits from the following two sources:
\begin{enumerate}[label={(\alph*)}]
\item ``private'' random bits from the Solver, and 

\item the hash of the block mined immediately after the block containing the Solver's solution.
\end{enumerate}
By \emph{hash of the block}, or \emph{block hash}, we more precisely mean the hash of the block's \emph{header}, a roughly 200-byte piece of data that contains the timestamp, nonce, previous block hash, and Merkle root hash of the transactions occurring in the block~\cite{ewhite}.  Source~(b) achieves properties~1.\ and 2.\ above, and source~(a) achieves property~3.  

\truebit generates Solvers' ``private'' random bits using a method reminiscent of RANDAO's random generator mechanism \cite{RANDAO}.  Before participating in a \truebit contract, a Solver must privately generate a string of random bits~$r$ and publish its hash on the blockchain.  This action commits the Solver to using~$r$ in the protocol without revealing the value~$r$ to others.

The Solver establishes property~2.\ above by committing both a hash of a ``correct'' solution and a hash of an ``incorrect'' solution prior to the broadcast of the block hash from item~(b).  At the time that this block hash is revealed, only the Solver, who has access to the value $r$, knows whether or not a forced error is in effect.  He publicly designates one of his two hashed solutions for evaluation, however potential Verifiers do not know \emph{a prioiri} whether the Solver intended his solution to be ``correct'' or ``incorrect.''  Only after the timeout for challenges has elapsed do the Verifiers witness the Solver's private random bits in the clear and learn whether a forced error was in effect.  In case the Solver's solution is challenged, the Solver must reveal $r$ as well as his second ``correct'' solution in order to prove that the error was forced and avoid a penalty (see Section~\ref{sec:lhfruit}).  Thus, in the case of a forced error, the Task Giver still obtains a correct solution.  In case of an unforced error, i.e.\ when the Solver evidently fails to provide a correct solution, the Task Giver receives a full refund of her task reward and taxes (see Section~\ref{sec:deposit} and Section~\ref{sec:truebit protocol}).

Although in theory one could securely run this part of the protocol without hashing the Solver's solutions, hashing makes it easier for the Solver to provide a convincing ``incorrect'' solution which appears, upon casual inspection, indistinguishable from a ``correct'' one.  The Solver can effectively use any randomly selected ``incorrect'' solution during a forced error because he never has to reveal its preimage.

\subsection{Solver and Verifier election} \label{sec:solverselection}

The system effectively chooses Solvers by lottery.  When a task is announced, Solvers broadcast their interest in solving it to the Referees in the form of Ethereum transactions.  Referees, or more specifically miners, choose one of these transactions to include the next block, thereby electing the Solver for that given task.  In case the miner who chooses the transactions for the current block happens to also hold a lottery ticket, he may bias his chances of winning the lottery.  This bias does not affect the security of \truebit, however, since the Solver must still provide a correct solution.

In contrast to the Solver selection, \truebit does not impose a limit on the number of Verifiers, and we describe in Section~\ref{sec:deeppocket} how multiple Verifiers who participate in a task must split their rewards after a successful challenge.  Due to the computational work involved in verification, rational Verifiers will only verify tasks in case they expect to receive adequate compensation for their work (assumption~\textsc{(ii)}, Section~\ref{sec:netass}).  Thus the number of Verifiers verifying each task will remain low due to the balance between computation costs and incentives (see Section~\ref{sec:trifecta} and Section~\ref{sec:deeppocket}).

\subsection{Protocol overview} \label{sec:truebit protocol}

In this section we present an overview of the \truebit protocol.  We will discuss how our chosen parameters enable system security in Section~\ref{sec:defenses}.  Throughout the description below, Solvers and Verifiers who deviate from protocol by broadcasting malformed data or failing to respond within timeout bounds forfeit their security deposits to the jackpot repository.
\begin{description}
\item[Preprocessing steps.]
The following must be done prior to commencing \truebit operation:
\begin{enumerate}
\item A substantial jackpot repository must be established for the\linebreak \truebit contract prior to commencement of any interactions  (See Section \ref{sec:jackpot} for details).

\item Solvers and Verifiers that wish to participate must commit deposits to the \truebit smart contract (See Section~\ref{sec:deposit}).  The deposits may be placed at any time during the protocol so long as the underlying contract has sufficient time to confirm receipt of the funds.

\item Solvers must also generate private random bits and commit their respective hashes to the blockchain.  We denote the private random bits of the unique \solver selected to perform the \task below by~$r$ (see Section~\ref{sec:randombits} for more details).

\item A universal tax rate $T$ must be established (see Section~\ref{sec:tax}).
\end{enumerate}

\item[Main algorithm.]  The protocol steps are as follows.
\begin{enumerate}
\item A Task Giver provides the following:
\begin{enumerate}
\item a computational \task,
\item the \timeout for accepting bids, performing the computation, and waiting for a challenge.  In the protocol below, we do not distinguish between these various timeouts with distinct notation, and we colloquially use ``\timeout'' to refer to both events and lengths of time.  In all cases, \timeout must be long enough to avoid microforks during which Referees temporarily disagree on current parameters.
\item a \reward for a correct output, which must be at least the cash equivalent of the task difficulty~$d$ based on \timeout (see Section~\ref{sec:cashequivalence}), plus a total tax of $T \cdot d$.  The \reward is held in escrow by the \truebit contract while the taxes are immediately deposited into the jackpot repository.
\item the minimum deposit, \mindeposit, needed to participate as a Solver or Verifier (see Section~\ref{sec:deposit}, Section~\ref{sec:sybil}, Section~\ref{sec:trifecta}, and Section~\ref{sec:deeppocket}).

\end{enumerate}

\item Solvers who have the requisite \mindeposit and random bits can bid to take on the task until the bidding \timeout.  At most one Solver is selected (Section~\ref{sec:solverselection}), and if no Solver takes on the task in the allotted time, the task request is canceled and the Task Giver receives a full refund.

\item The \solver privately computes \task.  In case of a \timeout, \solver forfeits his deposit to the jackpot repository and the protocol terminates.  
\begin{enumerate}
\item \solver commits two distinct hashes to the blockchain, thereby committing both a ``correct'' and an ``incorrect'' solution.

\item The hash of the next mined block is revealed, and then \solver knows whether or not there is a forced error (see Section~\ref{sec:randombits}). 

\item \solver designates one of the two hashes as the hash of his \solution.
\end{enumerate}

\item Verifiers who have posted \mindeposit can challenge (the hash of) \solution until \timeout.  Prior to \timeout, the Verifier must broadcast the hash of an even integer to the blockchain in order to commit to a challenge.  Hashing an odd number in case of no challenge is optional and may be used to camouflage real challenges from other Verifiers (see Section~\ref{sec:deeppocket}).  After \timeout, the Verifier broadcasts to the blockchain this hashed number in the clear to reveal her action.
\begin{enumerate}
\item If no Verifier challenges \solution, then 
\begin{enumerate}
\item \solver reveals $r$ to prove that there was no forced error (i.e.\ the criteria in Step~4(b)i\ below fails), 
\item \solver reveals \solution in the clear on the blockchain,
\item \solver receives the task \reward, and then
\item the protocol terminates.
\end{enumerate}

\item Otherwise, some \verifier challenges \solution.
\begin{enumerate}
\item  \solver reveals his private random string~$r$ on the blockchain, and Referees check it against his commitment from preprocessing step~3.  If the hash of the concatenation of $r$ and the block hash following the \solution announcement from~3(b) is small (as determined by the forced error rate, see Section~\ref{sec:jackpot}), then a forced error is in effect (see Section~\ref{sec:randombits}).

\item If the value $r$ reveals that \solver's error was forced, then \solver must reveal his secondary solution in the clear (see Section~\ref{sec:lhfruit}).
\begin{enumerate}
\item If no Verifer challenges \solver's secondary solution solution before \timeout, then \verifier wins a fraction the maximum jackpot amount $J$, scaled for task difficulty.  In case of challenges from multiple Verifiers, the jackpot is split among them.  In more detail, if there are $k$ distinct challenges, then each participating Verifier receives $J/2^{k-1}$ times the ratio of the task's difficulty to the system's maximum task difficulty (See Sections~\ref{sec:jackpot},~\ref{sec:deeppocket}, and~\ref{sec:lhfruit} for further discussion).

\item Otherwise the \solver must play the verification game with the challenging Verifier(s).  Verifier penalties, Verifier rewards, \solver penalties, and refunds to the Task Giver here are the same as described in the next step.  In case the \solver successfully defends himself against all challenges, however, then the jackpot payouts proceed as in Step~4(b)i above.
\end{enumerate}

\item Otherwise the error was not forced.  \solver reveals \linebreak \solution in the clear, and then \solver and \verifier must play a verification game  (Section~\ref{sec:verificationgame}).  In case of challenges from multiple Verifiers, the steps below are repeated until either some Verifier wins a challenge or \solver defeats all Verifier challenges.  The \solver collects \reward only once he wins all of the challenges, and if this does not happen, then the Task Giver receives a refund on his \reward, and tax payments are reimbursed through \solver's deposit as described below.
\begin{enumerate}
\item If \solver wins, then \verifier forfeits half of her deposit to the jackpot repository and the other half to the \solver, where the latter suffices to compensate the \solver for playing the verification game (see Section~\ref{sec:deposit}).

\item Otherwise \verifier wins.  \solver pays at most half of his deposit to the Verifier(s), according to the distribution scheme in Step~4(b)ii above, pays back the Task Giver's taxes out of this amount, and forfeits the remaining funds to the jackpot repository (see Section~\ref{sec:sybil} and Section~\ref{sec:trifecta}).
\end{enumerate}
\end{enumerate}
\end{enumerate}
\end{enumerate}
\vspace{-1ex}
\item[End of protocol.]
\end{description}

Note that \solver does not receive \reward when a forced error occurs in Step 4(b)ii.  In such situations, we expect the \solver to challenge his own ``mistake'' and receive a jackpot payout much greater than \reward.  \truebit makes this payout automatic since the \solver would lose the verification game anyway if he and \verifier were to play it out.

\subsection{Sanity check}

The table on the next page recaps the parameters used in the \truebit protocol and hints at how to estimate their exact values.  Each parameter is either a fixed constant, an input from a participant or something already on the blockchain, or can be computed from the elements which appear above it in the table.  By inspecting the list below from top to bottom, one can confirm that all notions are well-defined.
\begin{table}
\begin{tabular}{p{5cm}|p{6.5cm}}
\textbf{parameter} & \textbf{dependency} \\
\hline
dispute layer parameters \linebreak\phantom{qu} $p$, $c$, and $\sigma$   (Section~\ref{sec:amplification layer}) & - fixed constants.\\
tax rate \phantom{blah blah blah blah}\linebreak\phantom{qu}(Section~\ref{sec:tax}, Section~\ref{sec:trifecta}) & - fixed constant (500\% -- 5000\%).  \\
forced error rate \phantom{blah blah}\linebreak\phantom{qu} (Section~\ref{sec:jackpot})& - fixed constant (1/1000).\\
maximum jackpot payout \linebreak\phantom{qu} (Section~\ref{sec:tax})& - $1/3$ of current jackpot repository.\\
cash equivalent of CPU cycles \linebreak\phantom{qu}(Section~\ref{sec:cashequivalence})& - based on external markets.\\
maximum task difficulty \linebreak\phantom{qu}(Section~\ref{sec:jackpot}) & - maximum jackpot payout divided by cash equivalent of a CPU cycle.\\
\hline
task & Parameters in this box are chosen by the Task Giver with minimums as described below:\\
timeouts  (Section~\ref{sec:truebit protocol})& - long enough to avoid microforks. \\
task difficulty & - determined by timeouts. \\
Solver reward (Section~\ref{sec:cashequivalence}) & - no less than the cash equivalent of the task difficulty.\\
expected jackpot payout \linebreak\phantom{qu} (Section~\ref{sec:jackpot}, Section~\ref{sec:trifecta}, \linebreak\phantom{ququ}Section~\ref{sec:lhfruit}) & - cash equivalent of task difficulty, and number of active Verifier deposits.\\
Solver \& Verifier deposits \linebreak\phantom{qu} (Section~\ref{sec:deposit}, Section~\ref{sec:sybil}, \linebreak\phantom{ququ}Section~\ref{sec:trifecta}, Section~\ref{sec:deeppocket})  &  - more than the cost of verification game plus the expected jackpot payout.  Also depends on tax rate.\\
\hline
actual number of Verifiers \linebreak\phantom{qu} (Section~\ref{sec:solverselection})& - as many join (incentives limit overparticipation).\\
jackpot payout for challenging \linebreak\phantom{qu} a forced error (Section~\ref{sec:deeppocket}) & - based on maximum jackpot payout, actual number of verifiers, and ratio of task difficulty to maximum task difficulty.\\
Payout for detecting an \phantom{qu}unforced error \phantom {blah blah}\linebreak\phantom{qu}(Section~\ref{sec:sybil},  Section~\ref{sec:trifecta})&  - at most half of Solver deposit is split among all Verifiers  (rest goes to jackpot repository).\\
\end{tabular}
\captionof*{table}{\textbf{Table}: Relations between parameters used in \truebit.}
\end{table}

\section{Defenses} \label{sec:defenses}

We now analyze the security of the incentive layer from Section~\ref{sec:incentive layer}.  \truebit's security relies on the presence of at least one Verifier to check each task performed.  In this section we show that available incentives suffice to guarantee the existence of such Verifier(s) according to the network assumptions in Section~\ref{sec:netass} and the attacker model from Section~\ref{sec:attackermodel}.  \truebit defends against shortcuts which could divert verification incentives to non-verifying parties.  In particular, we show that \truebit resists Sybil attacks, collusion pools, opportunistic attacks related to jackpot payoffs, and certain external threats.

\subsection{Pairwise Sybil attacks} \label{sec:sybil}
In a Sybil attack, an adversary assumes multiple identities on the network in order to execute an exploit.  Identities on \truebit include Task Givers, Solvers, Verifiers, Judges, and Referees.  By our assumption (Section~\ref{sec:attackermodel}), Judges and Referees always function as intended, and we provide additional justification for this axiom here.  In this subsection, we consider all sets of pairs among the remaining identity types and show that pairwise cooperation does not harm operation of the system.  While parties can freely join and leave \truebit, each identity must make a deposit in order to participate.  This deposit alone is a general deterrent against Sybil attacks, however as we shall see, multiple identities do not provide much cheating advantage.

\paragraph{Judges and Referees.}  Recall that Ethereum miners play the roles of Judges and Referees in \truebit.  Our analyses in Section~\ref{sec:vgruntime} and Section~\ref{sec:truebit protocol} show that the work done by these parties is small, and hence they are not vulnerable to a Verifier's Dilemma (Section~\ref{sec:verifiersdilemma}).   Nakamoto consensus \cite{Nak09} therefore ensures that miners will not post bogus transactions, i.e.\ enforce rules incorrectly, lest other miners reject their block and cost them a block reward.  Therefore the only threat from miners is block withholding, specifically with regard to random number generator bias (see Section~\ref{sec:randombits}).

While in theory miners could discard blocks, along with their mining reward and associated random bits for \truebit, in practice this may never happen.  A miner who drops a block must expect in exchange some income greater than the usual mining reward in order to make this action worthwhile.  If such income were to come as a bribe from a \truebit participant, it would have to come from a Solver since only the Solver, who has unique access to his own private random bits, knows how to bias miners' random bits towards or away from a forced error.  In short, miners cannot disturb randomness in \truebit without a tip-off from a Solver.  The Solver, however, has no reason to bias against a forced error because this would prevent him from challenging his own answer and winning the jackpot, or at the very least would invite others to share it with him, thereby decreasing the total jackpot payout (see Section~\ref{sec:deeppocket}).  Moreover, the Solver is unlikely to succeed in biasing towards a forced error since miners have little control over their own block hashes.  This ``one-sided'' effect of block withholding, which can lock away jackpot funds but (almost) never release them, makes block hashes a safe source of randomness in \truebit.  Hence the Solver's potential reward for challenging his own solution under a forced error is not merely an artifact of \truebit's incentive structure --- it guarantees unbiased Referees.

\paragraph{Task Giver and Solver.}
A Task Giver who also acts as a Solver does not benefit from solving her own task.  One idea for an attack vector would be to create a task such that the Solver's private random bits force an error for the given task.  Since the Task Giver cannot predict the random bits from the block hash at the time the task is created (Section~\ref{sec:randombits}, part~(b)), the Task Giver cannot create such a task.  Moreover, the Task Giver cannot afford to flood \truebit with useless tasks and solve them herself in the hopes of eventually, by chance, encountering a forced error.  The Task Giver would pay more taxes to the jackpot repository than she would expect to win from the jackpot payout (Section~\ref{sec:tax}), even taking into consideration any rewards she might win by correctly solving her own tasks.  Her losses are amplified through payments to other Solvers, Verifiers, and Judges who choose to participate in the protocol.

\paragraph{Solver and Verifier.}
The Solver's burned deposit always exceeds the Verifier's income for successfully challenging an unforced error (see Section~\ref{sec:deposit}).  Hence the Solver has no incentive to challenge himself as a Verifier in such cases.  Similarly, a Verifier can only lose her deposit by posing bogus challenges.  In certain situations it is conceivable that a Solver--Verifier pair could benefit from submitting a false solution and then challenging it due to temporal constraints external to \truebit itself (as mentioned in Section~\ref{sec:deposit}), and in such cases the Task Giver must determine the necessary deposits to deter such actions.

In the case of an forced error, we expect the Solver will challenge himself in order to win the jackpot.  Nevertheless, \truebit's tax rate (Section~\ref{sec:tax}), and hence its jackpot payout (Section~\ref{sec:jackpot}), suffices to incentivize an independent Verifier to also check the solution.  As we shall see in Section~\ref{sec:deeppocket}, the Solver lacks incentive to claim the same jackpot more than once, hence the Solver's self-verification does not spoil other Verifiers' motivation to participate.

\paragraph{Task Giver and Verifier.}
A Task giver can certainly verify the solution she receives. This checking can only improve the reliability of the system!  We cannot, however, assume that the Task Giver will always perform such checks due to the Task Giver's possible computational and financial resource constraints or lack of motivation to involve herself in the details of a \truebit contract.

\subsection{The trifecta} \label{sec:trifecta}
Up until this point, we have implicitly assumed that the attacker's goal was to extract money from the jackpot without performing verification.  However there is another possibility: the attacker wishes to get a bogus solution accepted onto the blockchain.  No penalty scheme can deter all such attacks since \truebit, as a closed system, has no way to estimate the true economic impact of a bogus solution on Ethereum's blockchain.  Hence we now consider scenarios in which a party is willing to sacrifice a jackpot payout (or more) in order to skew computational results.

\paragraph{Scaring off Verifiers.}
Suppose that a Task Giver wishes to obtain an incorrect solution for a task.  Posing as a Solver, she gives a wrong solution to her own task.  Now the only thing that could stand in the way of her success would be a pesky Verifier.  In order to dodge this obstacle, the attacker poses as a regular Verifier prior to posting her malicious task.  She checks every task that comes along, until eventually he encounters a forced error.  At this point, the attacker challenges not once but a hundred times, which according to Step~4(b)ii in Section~\ref{sec:truebit protocol} essentially wipes out the jackpot payout for herself and any other Verifiers that also challenged this solution.  The upshot is that legitimate Verifiers no longer wish to participate because \truebit failed to deliver their well-deserved jackpot payout, and more to the point, they have no way to estimate their income from future jackpots.  So when the attacker finally broadcasts her malicious task, legitimate Verifiers simply ignore it.

As Verifier participation, and hence security, depends entirely on the expected value of jackpot payouts (assumption~\text{(ii)} in Section~\ref{sec:netass}), \truebit must ensure sufficient expected value for these payouts.  We therefore offer the following default strategy for all Verifiers.  New tasks are submitted to \truebit at various and overlapping times.  A prospective Verifier who wishes to earn rewards for verification should count the number of active Verifier deposits and compare this to the average number of simultaneous tasks occurring on the network.  If the tax rate and task flow profitably support an additional Verifier, then the prospective Verifier submits a deposit, randomly chooses an active task, and begins verifying.  As soon as she finishes verifying this task, she randomly selects another active task and begins verifying that next one.  Thus a Verifier with either more powerful hardware or better algorithms can gain a market advantage in \truebit.  Such a Verifier can check more solutions per unit time and therefore increase her expected jackpot rewards per unit time.  Verifiers are free to leave the network at any time, for any reason.  If the current number of active Verifiers exceeds the number supported by the tax rate and task flow, then the probabilistic default strategy tells each Verifier to exit the system with the minimal probability that one would expect to restore equilibrium if all Verifiers were following the default strategy.

Verifiers who choose tasks to verify at random according to the default strategy, as opposed to changing their behavior based on guesses as to who-solved-what before, can reasonably estimate their compensation for participation.   When a significant number of Verifiers (not necessarily a majority) follow the default strategy, the adversary wishing to establish a bogus solution can neither predict which tasks will be verified, nor can he influence which tasks default strategy Verifiers will verify.  According to assumption~\text{(ii)} in Section~\ref{sec:netass}, there exists some bound on the adversary's financial means.  If not, he could simply flood the Verifier pool with deposits and begin verifying, which in turn would incentivize legitimate Verifiers to leave the system.  Therefore the adversary can only post a bogus solution and hope that, by chance, it gets ignored.  In this way, the default Verifier strategy spoils the scare-off strategy above.

Let us now estimate the chances that the adversary succeeds to post a bogus solution which all legitimate Verifiers fail to notice.  We make the worst-case assumption that the adversary Solver always gets elected to solve his own task.  Suppose that the tax rate (see Section~\ref{sec:tax}) supports at least 6 Verifiers per task, and that the adversary holds at most $1/6$ of all Verifier deposits.  Then effectively there are an average of 5 verifiers per task.  Suppose $n$ is the average number of simultaneous tasks on the network.  Then the number of Verifiers is approximately $5n$, and so the chance that none of them check a particular solution is the chance that each of them independently and randomly choose other tasks, or $[(n-1)/n]^{5n}$.  Therefore the probability that the adversary gets caught, regardless of $n$, is
\[
1 - \left(1 - \frac{1}{n}\right)^{5n} > 1 - e^{-5} > 99\%.
\]
The leftmost inequality follows from the standard limit definition for the Calculus constant~$e$.  We conclude that the adversary will most likely lose his Solver deposit by attempting such an attack.

By raising taxes further, we can make such an attack even more costly.  In this sense the tax rate, together with the minimums for Solver deposits, bounds the mandatory ``honesty'' of the network.  If bogus solutions never occur, then there would be no need for taxes.  On the other hand, higher taxes make it more expensive to cheat.  There is a tradeoff between overhead tax expenses for Task Givers, minimum deposits for Solvers, and security of computations.

\paragraph{High-stake tasks.}
For some high-stakes applications it is possible to entirely fall back on the dispute resolution layer of Section~\ref{sec:amplification layer}.  In a high-stakes situation, it is reasonable to expect that other parties who have a significant financial stake in the outcome of a given task will make an effort to verify the solution regardless of expected jackpot payouts. This motivation acts as a secondary deterrent against the ``Scaring off Verifiers'' attack above, as does the fact that the Verifier receives a fraction of the Solver's deposit in case of positive error detection (\truebit burns part of the Solver's deposit in order to avoid collusion between Solvers and Verifiers).

\subsection{Collusion pools} \label{sec:deeppocket}

In this section, we analyze the potential effects of pooling computational, informational, and financial resources among Verifiers and Solvers.  Participants in \truebit may voluntarily form a \emph{Cartel} which may in turn use external smart contracts to enforce cooperation among its (mutually distrusting) members, however, Solver and Verifier deposits in \truebit alone prevent many potentially harmful financial collusions designed to extract jackpot funds without exerting computational effort.  Secondly, carefully chosen jackpot payout amounts, together with Solver and Verifier incentives and private commitment channels, prevent financial gain from unintended signaling of private random bits or challenges.

\paragraph{Rich but powerless.}
First, consider a Cartel of Solvers with limited CPU bandwidth but deep pockets which volunteers to solve each and every task that comes along.  The Cartel makes no attempt to provide correct solutions to any of these tasks, but instead intends to absorb lost Solver deposits until a forced error comes along, at which point it splits the jackpot among its members.  By construction (see Section~\ref{sec:deposit}), Solver deposits exceed the expected jackpot payout per task.  Therefore, in the long run, such a Cartel strategy loses money.  If members of the Cartel instead decide to start producing correct solutions, this does not harm security because that's what they are supposed to do anyway.  We remark that, also by construction (Section~\ref{sec:randombits}), a Solver cannot simply decide to accept only tasks which have forced errors because the Solver does not know whether or not a forced error is in effect until after he has committed his solution to the blockchain.

Similarly, a Cartel which challenges every single task in the hopes of eventually obtaining a jackpot sustains losses in the long run due to lost Verifier deposits.

\paragraph{A flood of Challengers.}
Timely information about forced errors has economic value.  While Solvers can earn jackpots by challenging their own forced errors, they risk sharing jackpot payouts by divulging information about such errors.  It follows that Solvers have incentive to keep their knowledge of forced errors private and their fake solutions convincing.  Information about active challenges also has economic value because a challenge posted to the blockchain could be a tip-off about a forced error.  For this reason, Verifiers have incentive to announce ``fake'' challenges rather than remaining silent on tasks in which they don't detect errors (See Step~5.\ in Section~\ref{sec:truebit protocol}).  ``Fake'' challenges serve as noise to mask real challenges and protect Verifiers' jackpot revenue from potential copycats who might challenge just because they see someone else doing it.

If the jackpot for a given task were a fixed amount and equally divided among challenging participants, then one could potentially flood the set of participants with aliases to obtain more than one's share of the jackpot.  A Cartel in which members inform each other about forced errors could therefore have the economic means to flood the network with challenges and monopolize jackpot payouts.  This in turn would dilute jackpots for legitimate Verifiers which would, in turn, degrade Verifier incentives for participation.  While such a Cartel might successfully detect forced errors, it might also cause tasks with unforced errors to go unverified.

In order to dissuade a Cartel from flooding the network with harmful challenges, we must reduce the total jackpot payout based on the number of challenges that occur.  The total jackpot paid out on a forced error with $k$ challengers must not exceed
\begin{equation} \label{eqn:jackpotpayoff}
\frac{J}{2^{k-1}}.
\end{equation}
The reason is as follows.  We want a Challenger's net reward to decrease if she superfluously clones herself via the Cartel.  Assume that there are $n$ legitimate challenges plus a Cartel which initially contributes $k$ challenges to the pool and is considering adding another alias.  We want the per challenge jackpot distribution among $n+k$ challenges to be less than it would be among $n+k+1$ challenges regardless of $k$, and $n$.  Let $J_i$ denote the total reward paid out when there are exactly~$i$ challenges.  Then we want
\[
\frac{k}{n} \cdot J_{n+k} > \frac{k+1}{n+1} \cdot J_{n+k+1},
\]
or
\[
J_{n+k+1} < J_{n+k} \cdot \underbrace{\frac{n+1}{n}}_{>1 \text{ for all } n \geq 0} \cdot \underbrace{\frac{k}{k+1}}_{\geq 1/2 \text{ for all } k \geq 1}.
\]
Thus it suffices to set $J_{n+k+1} = J_{n+k}/2$, or by induction $J_{k} \leq J_1/{2^{k-1}}$.  In fact we cannot do better than this because the case $k=1$ asymptotically forces a matching bound.

The upper bound in \eqref{eqn:jackpotpayoff} does not take into consideration the Verifier's cost of playing the verification game.  In the case where the Solver's solution has been found to have an error through one of the challenges, there is no need to repeat the verification game with the other challenges.  The Verifier who actually plays the verification game with the Solver receives a bonus from the Solver's deposit to compensate her for the CPU cycles spent during the challenge.

\subsection{On low-hanging fruit} \label{sec:lhfruit}
We analyze the security implications of Verifiers and Solvers who opportunistically disappear for hard tasks and reappear for easy ones, or who take advantage of solution discrepancies.

\paragraph{Easy winners.} If the jackpot for all tasks were equal, rational Verifiers might choose to verify only simple tasks and ignore complex ones.  For this reason, the jackpot for each task scales proportionally with the task's complexity (Step~5.\ in Section~\ref{sec:truebit protocol}).  Scaling ensures that a Verifier's expected jackpot payout per CPU cycle remains constant across all tasks, and it equally incentivizes Verifiers to inspect simple and complex tasks. Solvers always appear to perform tasks since the minimum task reward suffices to compensate them for their work (see Section~\ref{sec:truebit protocol} and  assumption~\textsc{(ii)} in Section~\ref{sec:netass}, ). 

\paragraph{Multiple solvers.}  For security reasons, \truebit explicitly does not allow Task Givers to hire redundant Solvers for a single task (Main Algorithm, Step~2., Section~\ref{sec:truebit protocol}).  Suppose that two Solvers provided solutions to a single task and exactly one of them receives a forced error.  Any Observer who notices a difference in the two solutions on the blockchain could, without verifying anything, challenge both solutions.  By playing two verifications and sacrificing one deposit, such an Observer could potentially win a jackpot at negligible cost, thereby degrading the overall incentives for verification.

\paragraph{Forced error in disguise.} 
What happens if a Solver has a forced error but gives a correct solution anyway?  The Solver could then challenge his own solution while other Verifiers ignore it, resulting in both the Solver receiving a bigger share of the jackpot as well as damage to the system's verification mechanism.  Therefore, when the Solver reveals that a forced error was in effect, the protocol also forces him to reveal his committed ``correct'' solution which must be distinct from the ``incorrect'' solution that he showed at first.  If the second solution revealed by the Solver is correct and distinct from his first solution, then by uniqueness of deterministic processes, the first solution must have been incorrect (as desired).  Verifiers have an opportunity to challenge this second solution.  If an error is detected in it via a verification game, the Verifier(s) split the Solver's deposit rather than the jackpot payout according to Step~4(b)ii in Section~\ref{sec:truebit protocol}.  Since the Verifier(s) presumably already performed the task themselves when checking the first solution, no additional CPU cycles are required to do this second check.  As the Solver loses both a jackpot payout and a deposit by following this course, by assumption~\textsc{(ii)} in Section~\ref{sec:attackermodel}, the situation described in this paragraph should never arise.

\subsection{A cash equivalence problem} \label{sec:cashequivalence}

Consider the following scenario.  A Task Giver wishes to get a bogus solution onto the blockchain.  He offers a minimal reward for a difficult task so as to ensure that no legitimate Solvers or Verifiers volunteer to inspect it.  Acting as a Solver, she then provides a bogus solution, and the system accepts her solution because no one bothers to check it.  It follows that for security purposes, the system must require that Task Givers compensate fairly based on the difficulty of the task.  But how can \truebit estimate the true cost of executing a given task in its native currency?

While one could potentially establish a long-term lower bound on the cost of a CPU cycle relative to a stable currency like the US dollar or Euro, calculating a lower bound relative to the value of a cryptocurrency token is another matter.  Cryptocurrencies are extremely volatile.  Moreover, \truebit lives on the blockchain and does not have access to a newspaper with current exchange rates.  

In the first iteration of \truebit, we will manually update the internal cash equivalent of a CPU cycle based on a live feed (e.g.\ \cite{coinmarketcap,oraclize,realitykeys}).  Ultimately, however, we would like to input these prices in a decentralized way without relying on a third-party.  Later versions of the protocol may make use of Augur \cite{augur}, a decentralized prediction market which ports outside information sources onto the Ethereum blockchain.  As of this writing, Augur is currently in beta testing.  Alternatively, we may build an independent blockchain for reporting prices whose ``transactions'' consist of exchange rate updates.

\section{Implementation} \label{sec:implementation}

Formally, \truebit is a smart contract in Ethereum which uses Ethereum's existing smart contract framework to bootstrap the creation of computationally-intensive \truebit contracts.  Tasks in \truebit take the form of C, C++, or Rust
code, but the user must pass this code as input to the Google Lanai architecture~\cite{lanai} (discussed below) prior to submission to \truebit.  This latter step guarantees consistency of simulated architecture and allows Judges to adjudicate fairly.

\paragraph{Google's Lanai interpreter.} \label{sec:lanai}
The theoretical framework for the verification game requires a fixed computer architecture to be used for all verification tasks.
In \cite{CRR13}, the authors used the regular Intel X86 architecture for this purpose, but we believe that this architecture is far too
complicated to be used in \truebit.  All participants of the verification game, including the Judges, have to simulate the behavior of the entire architecture. Even a slight difference in these implementations could result in catastrophic loss of funds for one of the parties. Moreover, simpler architecture costs less to simulate on the blockchain, and because of its simplicity and the fact that there now exists an efficient and trusted LLVM compiler writer, we have chosen to use Google's Lanai architecture~\cite{lanai}.  The full Google Lanai interpreter will be available as a smart contract in Ethereum so that it can be used by Judges.  In all \truebit contracts, the Lanai bytecode will be authoritative.

\truebit's on-chain interpreter runs in Solidity.  For efficiency reasons, tasks will not be run using the interpreters except in case of dispute.  In general, users will run tasks written in native programming languages and running on regular hardware.  In the majority of cases where tasks do not require the Lanai interpreter, Solvers and Verifiers can privately optimize implementation of task executions and potentially gain a market
advantage over other participants.

\section{Applications} \label{sec:applications}

\truebit is more than just an outsourced computation system.  It is designed for use in trustless smart contracts.  We present some examples of possible use cases.

\subsection{Practical decentralized pooled mining} \label{sec:smartpool}

Mining rewards are extremely competitive.  A typical desktop computer might only mine on average one Bitcoin block every thousand years.  To reduce income variance, miners often join {\em mining pools} which share both computing resources and mining rewards.   Each pool has an \emph{operator} who distributes computing tasks and rewards to pool \emph{members}.  This centralization degrades the security of the system by giving the operator undue influence to censor transactions~\cite{LWTS16re}.  In extreme cases where an operator controls more than half of the network's hash rate, as has been the case with \texttt{DwarfPool}~\cite{dwarfpool} in Ethereum, \texttt{GHash.io}~\cite{ghashio} in Bitcoin, and could happen again with Bitmain's proposed gigantic mining center~\cite{bitmain}, the operator can even withdraw cleared transactions and double-spend money by way of a \emph{51\% attack}~\cite{KDF13}.

SmartPool \cite{smartpool} introduces mining pools whose operators are Ethereum smart contracts.  As decentralized pool operators, smart contracts have many virtues.  In addition to counteracting the censorship issues described in the previous paragraph, they can operate at low cost relative to centralized pools and do not rely on a social contract to ensure fairness.   Unlike other decentralized mining pools, which either have higher payout variance~\cite{p2pool} that negate incentives for joining the pool or require a change in the proof-of-work protocol~\cite{nonoutsourceable}, SmartPool offers low variance payouts and retrofits existing cryptocurrencies, and it can handle a huge number of participants with a wide range of computing power.

SmartPool's proposed Ethereum mining pool minimizes verification work by precomputing the 1~GB data sets needed to check Ethereum's proof-of-work~\cite{ethashseed,LWTS16re}.  While this shortcut may help in checking Ethereum's proof-of-work, not all cryptocurrencies have this lucky feature.  Fortunately \truebit contracts can check \emph{any} proof-of-work, which means that with \truebit, we can build a smart contract-based mining pool for \emph{any} Nakamoto consensus-based cryptocurrency.  At the time of this writing, for example, the effort required to check a Zcash proof-of-work \cite{zcashpow} appears to exceed Ethereum's \texttt{gasLimit} capacity by a factor of 50.  \truebit is an option for bringing this task within reach of Ethereum smart contracts.  Project Alchemy \cite{alchemy}, which aims to bring smart contract functionality to Zcash in the spirit of the Dogecoin--Ethereum bridge below, may also benefit from \truebit's ability to check Zcash's proof-of-work.

\subsection{Dogecoin--Ethereum bridge}

We can use \truebit to build a two-way peg between Dogecoin~\cite{dogecoin} and Ethereum, enabling Dogecoin users to move dogecoins between Dogecoin's blockchain and Ethereum's blockchain without transmitting currency through a third-party exchange and effectively adding smart contract functionality to Dogecoin.  The Dogecoin community maintains an active interest in such a bridge~\cite{dogether}, and current offers a prize of more than 6000~ETH for its construction~\cite{dogebounty1,dogebounty2}.

\truebit contracts can check Dogecoin's Scrypt-based proof-of-work \linebreak whereas traditional Ethereum smart contracts cannot.  If Dogecoin were to enhance its scripting language with an operation indicating currency transfer to Ethereum addresses, a \truebit contract could then confirm such transfers.  The newly created dogecoin token could then be passed around Ethereum, and a final signed transaction in Ethereum could finally send the dogecoin back onto Dogecoin's blockchain, assuming that Dogecoin miners are willing to treat the Ethereum blockchain as authoritative for such transfers.

\subsection{Scalable transaction  throughput} \label{sec:scalable}

Building a system that can securely meet even the modest transaction volume demands of current Bitcoin users remains a formidable challenge \cite{CDEGJKMSSW16}.  Proposed Byzantine ``sharding''~\cite{AMNRS16,DSW16,KJGKGF16,LNBZGS16,Mic16,MXCSS16,PS16}, and miner-based ``serializing''~\cite{EGSR16, LSZ16} protocols exist which aim to distribute verification work, but here we take a simple and fundamentally different approach to decouple the two tasks which miners perform, namely
\begin{enumerate}
\item selecting which transactions to include in the blockchain, and

\item verifying that blockchain transactions are valid.
\end{enumerate}
Using \truebit, one can construct a verification protocol whose incentives guarantee that task~2 is correctly performed by off-chain Solvers and Verifiers (with some help from Judges and Referees), while miners continue to perform task~1.  In this way, complex transactions can securely reach the blockchain without overburdening miners.

\subsection{Towards a big data system} \label{sec:bigdata}
 
In order to perform as a truly scalable cloud computer, \truebit must have access to a scalable data storage system.  Ethereum's blockchain alone does not suffice as storing even moderate amounts of data directly on Ethereum's blockchain is prohibitively expensive.  \truebit can securely access and use portions of massive data sets so long as the data is stored somewhere publicly and permanently, for example in Swarm~\cite{swarm} or on another blockchain.  Parties who wish to rely on such data in a \truebit contract must be sure that Verifiers have access to the full data set.

To use \truebit on external data, one need only store a Merkle root of the massive data set on the blockchain and add non-deterministic steps in the verification game in which the Solver can ``guess'' the original data set represented by the Merkle root.  While Solvers and Verifiers must have access to the full data, Judges and Referees do not.   Indeed, if we modify the verification game so as to permit tasks for nondeterministic Turing machines, then the Solver can nondeterministically guess the certificate data as a step in the \truebit contract.   Only in cases of disputes would the Solver have to reveal external certificate data to the Judges via the blockchain.  In some applications, the Solver might even even be able to reveal to the Judges a privacy-preserving zkSNARK rather than the data itself.  zkSNARKs have the potential to enable privacy for many different kinds of systems on Ethereum~\cite{nutshell}.

While in theory \truebit's scalable protocol can process arbitrarily complex tasks, in practice the verification game is inefficient and therefore security of \truebit computations degrades as tasks reach the level of big data.  For big data applications, \truebit may not be able to rely on a one-size-fits-all verification game.  Therefore we anticipate optimizing the verification game for certain classes of tasks.  For example, Section~\ref{sec:verificationgame} gives an example of an efficient, specialized verification game for matrix multiplication.  In future versions of \truebit, Task Givers might broadcast not only tasks but also an indication of the corresponding verification game to be played in case of disputes.

\begin{remark}
We conclude with a caveat: \truebit may expose latent security vulnerabilities in the underlying Ethereum network as a result of new kinds of interactions between smart contracts and miners.  By allowing smart contracts to check proof-of-works, for example, \truebit may facilitate 38.2\% attacks~\cite{TJS16}.
\end{remark}

\paragraph{Acknowledgments.}  
We thank Vitalik Buterin and Vlad Zamfir for suggesting the use of forced errors in the \truebit protocol and Eli Bendersky for introducing us to Google Lanai.  We also thank Loi Luu and Julia Koch for useful discussions.

\appendix

\section{Addendum}
In the months following the initial release of this whitepaper, new work and feedback informed TrueBit's development roadmap and helped refine the protocol itself.  We discuss a few recent developments.

\subsection{Security patches}
A TrueBit adversary has either one of two goals:
\begin{enumerate}
\item get bogus computations onto the blockchain, or

\item  extract jackpot funds without performing verification.
\end{enumerate}
We investigate three attacks of the first type followed by two of the second.

\paragraph{Premature disclosure of random bits \textmd{(Zack Lawrence)}.}
A Solver can dissuade Verifier participation by publicly broadcasting his private random bits prior to the designated reveal time.  Verifiers then know immediately whether or not a forced error is in effect.  Since Verifiers expect to gain little from checking solutions without forced errors, they may decide not to verify, thereby offering opportunity to get bogus solutions onto the blockchain.

1protocol's \cite{1protocol} random number generator protocol, Arbit, solves this problem by instituting penalties for Solvers who prematurely reveal private random bits and rewarding users who report them.  When a user correctly reports a premature reveal to TrueBit, the following occurs.
\begin{enumerate}
\item The Solver's solutions are discarded, and a new Solver lottery takes place.  This re-incentivizes Verifiers to participate in the task while voiding the Solver's incentive to reveal private information.

\item Half of the Solver's deposit gets burned.  This makes the above attack expensive for the Solver.

\item The other half of the Solver's deposit goes to the party who reported the Solver's random bits.  This incentivizes Observers to report the Solver's prematurely revealed random bits.
\end{enumerate}

\paragraph{Incorrect secondary solution \textmd{(Sina Habibian and Harley Swick \cite{tbopenwiki})}.}
Suppose that a forced error is in effect and that the Solver submits two incorrect solutions.  When the Solver reveals his ``correct'' secondary solution in Step 4(b)ii of the protocol (Section~\ref{sec:truebit protocol}), Verifiers ignore it because there's no chance of a jackpot payout.  Indeed, the only ``reward'' for correctly challenging this secondary solution is to play a verification game.  Hence one of the Solver's bogus solutions ends up on the blockchain.

We eliminate the incorrect secondary solution vulnerability as follows.  Denote the Solver's two solutions by~$A$ and~$B$.  In the beginning of Step~4, rather than signaling for a challenge with the hash of an even integer, the Verifier hashes an integer whose value $\text{mod}~3$ the protocol interprets as follows:
\begin{center}
\begin{tabular}{ll}
$0 \mod 3$: & challenge solution~$A$,\\
$1 \mod 3$: & challenge solution~$B$,\\
$2 \mod 3$: & challenge both~$A$ and~$B$.
\end{tabular}
\end{center}
The Verifiers indicate their choice without knowing which of the two solutions the Solver puts forth as an answer.  The protocol hides this information from Verifiers via the following process.  The Solver commits to either solution $A$ or solution $B$ by hashing either $A$ or $B$ paired with his private random bits, where the Solver's private random bits serve as ``noise'' which prevent Verifiers from guessing which option the Solver chose.  The Solver has incentive not to share his private random bits due to the ``Premature disclosure of random bits'' patch above as well as the fact that the Solver risks reducing his jackpot share by exposing this private information.  Finally, once the timeout for challenges has passed, the Solver reveals his random bits in the clear, thereby indicating his commitment to either solution $A$ or solution $B$.  Challenges and forced errors then proceed as usual.  In case the protocol forces the Solver to reveal his second ``correct'' solution, Verifiers who earlier committed to a challenge against this second solution are obligated to play a verification game.  In this way, Verifiers catch errant secondary solutions just as they catch errant primary ones.

In case a forced error is \emph{not} in effect, broadcasting a pair of incorrect solutions poses a cost to the Solver in the form of a lost deposit.  Indeed Verifiers have proper incentives to check the Solver's primary answer.  Since forced errors occur rarely and unpredictably, the Solver expects to sacrifice several deposits in order to mount an ``incorrect secondary solution'' attack.  This loss offsets the Solver's eventual jackpot gain from challenging his own forced error solution.  We implicitly assume that the chance to win a jackpot sufficiently motivates a Verifier to challenge whenever a Solver submits a pair of incorrect solutions; any Verifier who challenges both submitted solutions must play a verification game. 

The fix above has a potentially useful side effect of publicly indicating how many Verifiers are monitoring a given task.  Indeed, a Verifier broadcasts one of the three commitment forms above if and only if she is paying attention.  The option to signal ``no challenge'' is no longer needed for camouflage because broadcasting challenges no longer indicates presence of a forced error.  Moreover, if the Solver were to submit two correct solutions, the smart contract could immediately recognize them as identical and penalize the Solver accordingly.

An adversary could potentially exploit the monitoring feature in the previous paragraph by broadcasting multiple challenge commitments from Sybil identities, thereby reducing the total payout in case of a forced error and discouraging other rational Verifiers from participating.  For this reason, the protocol must prevent Verifiers from revealing which task they are challenging until the final phase of the protocol.  Since each Verifier performs the given computational task without distraction from others' commitments, an adversary cannot deter Verifier participation via Sybil attack.

\paragraph{Program abort \textmd{(Yaron Velner)}.}
A task must explicitly specify an upper bound on the number of steps for which the verification game can run, and the Solver's solution should return an ``error'' if and only if the task computation exceeds this bound.  Indeed, the computation steps for a given task must form a deterministic sequence in order for Judges to determine their correctness.  Ideally, one would like to run tasks directly in a native language like C, C++, or Rust, however this approach requires accurate metering of the number of computation steps.  We can reduce the metering overhead by processing steps with a compiler rather than an interpreter.

\paragraph{Jackpot balloon attack \textmd{(Cl\'{e}ment Lesaege)}.}
In the ``flood of challengers'' attack (Section~\ref{sec:deeppocket}), a Cartel could artificially inflate the jackpot repository by repeatedly challenging a single forced error solution and thereby reduce jackpot payout.  Eventually, the Cartel could offset the cost of this attack by cashing in on an extra large jackpot.   This action could recruit membership for the Cartel at low cost by removing incentives for Verifiers who do not belong to the Cartel.  

We mitigate against the benefit of this attack by placing a fixed, absolute bound on the jackpot payout, regardless of the actual jackpot repository balance.  Extra revenue accumulated in the jackpot repository only becomes available for payout after a protocol upgrade. Note that the maximum jackpot amount does not necessarily pay out at each forced error; the actual payout depends on the difficulty of the task.

\paragraph{Incentivizing block withholding \textmd{(Cl\'{e}ment Lesaege)}.}
The attacker, who is a Solver, deploys a smart contract which pays money to miners who withhold (or intentionally uncle) blocks whose random seed fails to yield a forced error. See \cite{WTL16un} for possible implementations.  This attack is profitable for tasks in which the expected jackpot payout times the probability of a forced error exceeds a block reward.  Through repetitions, this attack could drain the jackpot repository.

Arbit's security mechanism, discussed in ``Premature disclosure of random bits'' above, defends against this attack.  In order to execute the present ``incentivizing block withholding'' attack, the Solver either has to reveal his bits publicly or has to provide them to miners through private channels.  If the Solver's penalty for a premature reveal exceeds the expected jackpot payout times the probability of a forced error, then this attack results in an expected net loss.

\subsection{The TrueBit Virtual Machine}

Section~\ref{sec:detailedprotcol} describes a verification game which operates with respect to a Turing machine.  In practice, however, no compiler exists which can transform C++ code into something as simple as Turing machine language.  In Section~\ref{sec:implementation}, we proposed Google Lanai as a practical compromise in the absence of Turing machine architecture.  Due to Lanai's complexity, the fact that Google controls its codebase, and that progress on its development appears to have slowed, we have since migrated development away from Google Lanai.

In addition to executing exactly the same computation steps regardless of hardware configuration, the complier architecture underlying the verification game, or \emph{TrueBit Virtual Machine (TVM)}, must satisfy the following \emph{simplicity properties}.
\begin{enumerate}
\item A single computation step on the TVM runs comfortably within Ethereum's gas limit, and

\item the space required to describe a TVM state change fits inside a single Ethereum transaction.
\end{enumerate}
WebAssembly architecture~\cite{wasm} comes sufficiently close to matching these properties so as to make TVM execution practical today.  WebAssembly has become increasingly ready-to-use due to contributions and testing by Apple, Google, Microsoft, and Mozilla \cite{HRSTHGWZB17}.  Several cryptocurrency projects have begun to develop on  WebAssembly, including Dfinity, eWASM, and Parity, due to the platform's machine independence and relative simplicity.

The TVM consists of two parts:
\begin{enumerate}
\item an off-chain \emph{interpreter} which enumerates a list of states for a given computation, and

\item an on-chain \emph{stepper} which, given a state, can compute the next state.
\end{enumerate}
Solvers and Challengers use the interpreter to create a Merklized list of states for a computation.  Once the Solver and Challenger have determined the first step at which they disagree, the Judges use the stepper to run this step and rule on the outcome of the verification game.

Since Native WebAssembly does not entirely satisfy the simplicity properties above,  the interpreter must either further compile input code into a simpler architecture, or it must divide each WebAssembly instruction into substeps.  In order to reduce the chances of inducing compiler error, the TVM follows the latter strategy.  The TVM's off-chain interpreter parses WebAssembly instructions into OCaml execution which in turn creates registers describing WebAssembly suboperations.  Each Ocaml-based sub-step only accesses one dynamic data structure at a time.

\subsection{Additional applications}

Finally, we mention a few more illustrative applications.

\paragraph{Video broadcasting.}  Livepeer \cite{livepeer} offers a new platform for decentralized, live streaming video.  Users can broadcast, watch, or get paid for performing the computationally intensive process of transcoding video into different codecs and formats.  TrueBit ensures that transcoders perform this work correctly, while Swarm \cite{swarm} guarantees that video data remains available to TrueBit during the necessary verification period.

\paragraph{Autonomous machine learning.}  McConaghy's ArtDAO \cite{artDAO} generates art, sells it, and then uses its revenue to improve its own code.  A TrueBit-based ArtDAO would allow a financed piece of code on the blockchain to access computational resources in such a way that no one can ``pull its plug.''  We may eventually see other blockchain-based machine learning applications, like computer vision, as well.

\paragraph{Data marketplace.}  Hedge fund Numerai \cite{numerai} crowdsources modeling \linebreak problems to data scientists and then executes trades based upon their work.  Numerai rewards successful models, however contributing data scientists must trust Numerai to both test their work and compensate them fairly.  TrueBit enables autonomous data markets. Open Mined~\cite{openmined}, paired with TrueBit, opens the possibility of trustless renumeration based on streaming flows of models and data.  The Ocean Protocol \cite{ocean}, which facilitates contribution and sharing of data, also requires a verification mechanism.

\paragraph{Staking and random numbers.} 1protocol \cite{1protocol} allows users who have either computing resources or capital, but not necessarily both, to participate as TrueBit Solvers and Verifiers by decoupling security deposits from work done.  In addition, 1protocol's Arbit protocol uses interactive verification to generate random numbers in a decentralized way.

\paragraph{Other applications.}  Please check the TrueBit website for other current ideas in progress! \url{https://truebit.io}.

\bibliographystyle{plain}
\bibliography{truebit}

\begin{thebibliography}{10}

\bibitem{1protocol}
1protocol.
\newblock \url{http://1protocol.com}.

\bibitem{augur}
Augur.
\newblock \url{https://www.augur.net/}.

\bibitem{bitmain}
Bitmain responds to controversy surrounding its upcoming 140,000 kw mining
  center.
\newblock
  \url{http://www.newsbtc.com/2016/11/04/bitmain-response-new-mining-center/}.

\bibitem{coinmarketcap}
Crypto-currency market capitalizations.
\newblock \url{http://coinmarketcap.com/}.

\bibitem{dogebounty1}
The {D}oge connection {B}ounty {D}ao is live and working.
\newblock
  \url{https://www.reddit.com/r/ethereum/comments/41ohhr/the_doge_connection_bounty_dao_is_live_and_working/}.

\bibitem{dogecoin}
Dogecoin.
\newblock \url{http://dogecoin.com/}.

\bibitem{dogebounty2}
{D}ogecoin--{E}thereum bounty smart contract.
\newblock
  \url{https://etherscan.io/address/0xdbf03b407c01e7cd3cbea99509d93f8dddc8c6fb}.

\bibitem{dogether}
Dogethereum.
\newblock \url{https://www.reddit.com/r/dogethereum/}.

\bibitem{dwarfpool}
Dwarfpool is now 50.5\%.
\newblock \url{http://forum.ethereum.org/discussion/5244
  /dwarfpool-is-now-50-5}.

\bibitem{ethashseed}
Ethash: defining the seed hash.
\newblock
  \url{https://github.com/ethereum/wiki/wiki/Ethash#defining-the-seed-hash}.

\bibitem{ethereum}
Ethereum.
\newblock \url{http://ethereum.org/}.

\bibitem{computationmarket}
{E}thereum {C}omputation {M}arket.
\newblock \url{http://www.ethereum-computation-market.com/}.

\bibitem{golem}
Golem.
\newblock \url{http://github.com/imapp-pl/golem/wiki/FAQ}.

\bibitem{golemwhite}
The {G}olem {P}roject: crowdfunding whitepaper.
\newblock \url{https://golem.network/doc/Golemwhitepaper.pdf}.

\bibitem{lanai}
{G}oogle {L}anai.
\newblock \url{http://llvm.org/docs/CompilerWriterInfo.html#lanai}.

\bibitem{ethdos1}
I thikn the attacker is this miner---today he made over \$50k.
\newblock
  \url{https://www.reddit.com/r/ethereum/comments/55xh2w/i_thikn_the_attacker_is_this_miner_today_he_made/}.

\bibitem{alchemy}
Introducing {P}roject {A}lchemy.
\newblock \url{https://z.cash/blog/project-alchemy.html}.

\bibitem{livepeer}
Livepeer.
\newblock \url{https://livepeer.org/}.

\bibitem{numerai}
Numerai.
\newblock \url{https://numer.ai}.

\bibitem{ocean}
{O}cean {P}rotocol.
\newblock \url{https://oceanprotocol.com/}.

\bibitem{openmined}
{O}pen {M}ined.
\newblock \url{https://openmined.org/}.

\bibitem{tbopenwiki}
Open problems.
\newblock
  \url{https://github.com/TrueBitFoundation/Developer-Resources/wiki/Open-Problems}.

\bibitem{oraclize}
Oraclize.
\newblock \url{http://www.oraclize.it/}.

\bibitem{p2pool}
{P2P}ool.
\newblock \url{http://p2pool.org/}.

\bibitem{RANDAO}
{RANDAO}.
\newblock \url{https://github.com/randao/randao}.

\bibitem{realitykeys}
{R}eality {K}eys.
\newblock \url{https://www.realitykeys.com/}.

\bibitem{smartpool}
Smart{P}ool.
\newblock \url{http://smartpool.io}.

\bibitem{swarm}
Swarm.
\newblock \url{http://swarm-gateways.net/}.

\bibitem{thedao}
The{ DAO}.
\newblock \url{http://daohub.org/}.

\bibitem{ghashio}
Warning: {G}hash.io is nearing 51\% -- leave the pool.
\newblock
  \url{http://www.cryptocoinsnews.com/warning-ghash-io-nearing-51-leave-pool/}.

\bibitem{wasm}
Web{A}ssembly.
\newblock \url{http://webassembly.org/}.

\bibitem{ewhite}
White paper.
\newblock \url{https://github.com/ethereum/wiki/wiki/White-Paper}.

\bibitem{zcashpow}
Why {E}quihash?
\newblock \url{https://z.cash/blog/why-equihash.html}.

\bibitem{zcash}
Zcash.
\newblock \url{https://z.cash/}.

\bibitem{july4fork}
Some miners generating invalid blocks.
\newblock \url{https://bitcoin.org/en/alert/2015-07-04-spv-mining}, July 2015.

\bibitem{AMNRS16}
Ittai Abraham, Dahlia Malkhi, Kartik Nayak, Ling Ren, and Alexander Spiegelman.
\newblock {S}olidus: An incentive-compatible cryptocurrency based on
  permissionless {B}yzantine consensus.
\newblock \url{https://arxiv.org/abs/1612.02916}, 2016.

\bibitem{BCGTV13}
Eli Ben-Sasson, Alessandro Chiesa, Daniel Genkin, Eran Tromer, and Madars
  Virza.
\newblock {SNARKs} for {C}: Verifying program executions succinctly and in zero
  knowledge.
\newblock In Ran Canetti and Juan~A. Garay, editors, {\em Advances in
  Cryptology -- CRYPTO 2013: 33rd Annual Cryptology Conference, Santa Barbara,
  CA, USA, August 18-22, 2013. Proceedings, Part II}, pages 90--108, Berlin,
  Heidelberg, 2013. Springer Berlin Heidelberg.

\bibitem{BCPR13}
Nir Bitansky, Ran Canetti, Omer Paneth, and Alon Rosen.
\newblock Indistinguishability obfuscation vs. auxiliary-input extractable
  functions: One must fall.
\newblock \url{https://eprint.iacr.org/2013/468.pdf}.

\bibitem{CRR11}
Ran Canetti, Ben Riva, and Guy~N. Rothblum.
\newblock Practical delegation of computation using multiple servers.
\newblock In {\em Proceedings of the 18th ACM Conference on Computer and
  Communications Security}, CCS '11, pages 445--454, New York, NY, USA, 2011.
  ACM.

\bibitem{CRR13}
Ran Canetti, Ben Riva, and Guy~N. Rothblum.
\newblock Refereed delegation of computation.
\newblock {\em Information and Computation}, 226:16 -- 36, 2013.
\newblock Special Issue: Information Security as a Resource.

\bibitem{CDEGJKMSSW16}
Kyle Croman, Christian Decker, Ittay Eyal, Adem~Efe Gencer, Ari Juels, Ahmed
  Kosba, Andrew Miller, Prateek Saxena, Elaine Shi, Emin G{\"u}n~Sirer, Dawn
  Song, and Roger Wattenhofer.
\newblock On scaling decentralized blockchains (a position paper).
\newblock In {\em Financial Cryptography and Data Security 2016 BITCOIN
  Workshop}, volume 9604 of {\em Lecture Notes in Computer Science}, pages
  106--125. Springer Berlin Heidelberg, February 2016.

\bibitem{DSW16}
Christian Decker, Jochen Seidel, and Roger Wattenhofer.
\newblock {B}itcoin meets strong consistency.
\newblock In {\em Proceedings of the 17th International Conference on
  Distributed Computing and Networking}, ICDCN '16, pages 13:1--13:10, New
  York, NY, USA, 2016. ACM.

\bibitem{EGSR16}
Ittay Eyal, Adem~Efe Gencer, Emin~Gun Sirer, and Robbert~Van Renesse.
\newblock {B}itcoin-{NG}: A scalable blockchain protocol.
\newblock In {\em 13th USENIX Symposium on Networked Systems Design and
  Implementation (NSDI 16)}, pages 45--59, Santa Clara, CA, March 2016. USENIX
  Association.

\bibitem{prubyreddit}
Tim Goddard.
\newblock {AdversariallyVerifiableMachine}.
\newblock
  \url{https://www.reddit.com/r/ethereum/comments/51qjz6/interactive_verification_of_c_programs/d7ey41n/}.

\bibitem{HRSTHGWZB17}
Andreas Haas, Andreas Rossberg, Derek~L. Schuff, Ben~L. Titzer, Michael Holman,
  Dan Gohman, Luke Wagner, Alon Zakai, and JF~Bastien.
\newblock Bringing the web up to speed with {W}eb{A}ssembly.
\newblock In {\em Proceedings of the 38th ACM SIGPLAN Conference on Programming
  Language Design and Implementation}, PLDI 2017, pages 185--200, New York, NY,
  USA, 2017. ACM.

\bibitem{JSST16}
Sanjay Jain, Prateek Saxena, Frank Stephan, and Jason Teutsch.
\newblock How to verify computation with a rational network.
\newblock \url{https://arxiv.org/abs/1606.05917}, June 2016.

\bibitem{KJGKGF16}
Eleftherios~Kokoris Kogias, Philipp Jovanovic, Nicolas Gailly, Ismail Khoffi,
  Linus Gasser, and Bryan Ford.
\newblock Enhancing {B}itcoin security and performance with strong consistency
  via collective signing.
\newblock In {\em 25th USENIX Security Symposium (USENIX Security 16)}, pages
  279--296, Austin, TX, 2016. USENIX Association.

\bibitem{KDF13}
Joshua~A. Kroll, Ian~C. Davey, and Edward~W. Felten.
\newblock {T}he economics of {B}itcoin mining, or {B}itcoin in the presence of
  adversaries.
\newblock
  \url{http://www.econinfosec.org/archive/weis2013/papers/KrollDaveyFeltenWEIS2013.pdf},
  June 2013.

\bibitem{LNBZGS16}
Loi Luu, Viswesh Narayanan, Chaodong Zheng, Kunal Baweja, Seth Gilbert, and
  Prateek Saxena.
\newblock A secure sharding protocol for open blockchains.
\newblock In {\em Proceedings of the 2016 ACM SIGSAC Conference on Computer and
  Communications Security}, CCS '16, pages 17--30, New York, NY, USA, 2016.
  ACM.

\bibitem{LTKS15}
Loi Luu, Jason Teutsch, Raghav Kulkarni, and Prateek Saxena.
\newblock Demystifying incentives in the consensus computer.
\newblock In {\em Proceedings of the 22nd ACM SIGSAC Conference on Computer and
  Communications Security (CCS 2015)}, pages 706--719, New York, NY, USA, 2015.
  ACM.

\bibitem{LWTS16re}
Loi Luu, Yaron Welner, Jason Teutsch, and Prateek Saxena.
\newblock {S}mart{P}ool: Practical decentralized pooled mining.
\newblock \url{http://smartpool.io/docs/smartpool.pdf}.

\bibitem{artDAO}
Trent McConaghy.
\newblock Wild, wooly {AI} {DAOs}.
\newblock \url{https://blog.bigchaindb.com/wild-wooly-ai-daos-d1719e040956}.

\bibitem{Mic16}
Silvio Micali.
\newblock {ALGORAND:} the efficient and democratic ledger.
\newblock \url{http://arxiv.org/abs/1607.01341}, 2016.

\bibitem{nonoutsourceable}
Andrew Miller, Ahmed Kosba, Jonathan Katz, and Elaine Shi.
\newblock Nonoutsourceable scratch-off puzzles to discourage {B}itcoin mining
  coalitions.
\newblock In {\em Proceedings of the 22Nd ACM SIGSAC Conference on Computer and
  Communications Security}, CCS '15, pages 680--691, New York, NY, USA, 2015.
  ACM.

\bibitem{MXCSS16}
Andrew Miller, Yu~Xia, Kyle Croman, Elaine Shi, and Dawn Song.
\newblock The honey badger of {BFT} protocols.
\newblock In {\em Proceedings of the 2016 ACM SIGSAC Conference on Computer and
  Communications Security}, CCS '16, pages 31--42, New York, NY, USA, 2016.
  ACM.

\bibitem{Nak09}
Satoshi Nakamoto.
\newblock Bitcoin {P2P} e-cash paper.
\newblock
  \url{http://www.mail-archive.com/cryptography@metzdowd.com/msg09959.html},
  November 2008.

\bibitem{PS16}
Rafael Pass and Elaine Shi.
\newblock Hybrid consensus: Efficient consensus in the permissionless model.
\newblock \url{https://eprint.iacr.org/2016/917.pdf}, 2016.

\bibitem{notsmartjudges}
Christian Reitwiessner.
\newblock From smart contracts to courts with not so smart judges.
\newblock
  \url{https://blog.ethereum.org/2016/02/17/smart-contracts-courts-not-smart-judges/}.

\bibitem{nutshell}
Christian Reitwiessner.
\newblock {zkSNARKs} in a nutshell.
\newblock \url{https://blog.ethereum.org/2016/12/05/zksnarks-in-a-nutshell/},
  Dececmber 2016.

\bibitem{LSZ16}
Yonatan Sompolinsky, Yoad Lewenberg, and Aviv Zohar.
\newblock {SPECTRE}: A fast and scalable cryptocurrency protocol.
\newblock \url{https://eprint.iacr.org/2016/1159.pdf}, 2016.

\bibitem{nick-szabo}
Nick Szabo.
\newblock The idea of smart contracts.
\newblock \url{http://szabo.best.vwh.net/smart_contracts_idea.html}, 1997.

\bibitem{TJS16}
Jason Teutsch, Sanjay Jain, and Prateek Saxena.
\newblock When cryptocurrencies mine their own business.
\newblock In {\em Financial Cryptography and Data Security: 20th International
  Conference (FC 2016) Christ Church, Barbados}, pages 499--514. Springer
  Berlin / Heidelberg, 2017.

\bibitem{HKZ14}
Jelle van~den Hooff, M.~Frans Kaashoek, and Nickolai Zeldovich.
\newblock Versum: Verifiable computations over large public logs.
\newblock In {\em Proceedings of the 2014 ACM SIGSAC Conference on Computer and
  Communications Security}, CCS '14, pages 1304--1316, New York, NY, USA, 2014.
  ACM.

\bibitem{WTL16un}
Yaron Velner, Jason Teutsch, and Loi Luu.
\newblock Smart contracts make {B}itcoin mining pools vulnerable.
\newblock To appear in \emph{4th Workshop on Bitcoin and Blockchain Research
  (BITCOIN 2017)}.

\bibitem{WB15}
Michael Walfish and Andrew~J. Blumberg.
\newblock Verifying computations without reexecuting them.
\newblock {\em Communications of the ACM}, 58(2):74--84, January 2015.

\end{thebibliography}

\end{document}